\documentclass[showpacs,preprint,superscriptaddress,showpacs,preprintnumbers,jcp]{revtex4-1}

\usepackage{graphicx}
\usepackage{dcolumn}
\usepackage[latin1]{inputenc}
\usepackage{latexsym}
\usepackage{amscd,amssymb,amsthm}
\usepackage{amsmath}
\usepackage{amsfonts}
\usepackage{color}
\usepackage{enumerate}

\newcommand{\V}{\mathbf v(\mathbf R)}
\newcommand{\shape}{\mathcal D_{S}}

\newcommand{\lsf}{\phi(\R,\tau)}

\begin{document}

\title{Maximum probability domains for the analysis of the microscopic structure of liquids}

\author{Federica Agostini$^a$, Giovanni Ciccotti$^{b,c}$, Andreas Savin$^{d,e}$, Rodolphe Vuilleumier$^{f,g}$\\
\vspace{6pt} $^{a}${\em{Max-Planck-Institute of Microstructure Physics, Weinberg 2, D-06120 Halle, Germany}}\\
\vspace{6pt} $^{b}${\em{Dipartimento di Fisica and CNISM Unit\`a 1, Universit\`a ``La Sapienza", Piazzale Aldo Moro 5, I-00185 Rome, Ital.}}\\
\vspace{6pt} $^{c}${\em{School of Physics, University College Dublin (UCD), Belfield Dublin 4, Ireland.}}\\
\vspace{6pt} $^{d}${\em{CNRS, UMR 7616, Laboratoire de Chimie Th\'eorique, F-75005 Paris, France.}}\\
\vspace{6pt} $^{e}${\em{Sorbonne Universit\'es, UPMC Univ Paris 06, UMR 7616, Laboratoire de Chimie Th\'eorique, F-75005 Paris, France.}}\\
\vspace{6pt} $^{f}${\em{UMR 8640 ENS-CNRS-UPMC, D\'epartement de Chimie, 24 rue Lhomond, \'Ecole Normale Sup\'erieure, 75005 Paris, France.}}\\
\vspace{6pt} $^{g}${\em{UPMC Universit\'e Paris 06, 4, Place Jussieu, 75005 Paris, France.}}\\
\vspace{6pt} $^{}$}
\date{\today}



\begin{abstract}
We introduce the concept of maximum probability domains (MPDs), developed in the context of the analysis of electronic densities, in the study of the 
microscopic spatial structures of liquids. The idea of locating a particle in a three dimensional region, by determining the domain where the 
probability of finding that, and only that, particle is maximum, gives an interesting characterization of the local structure of the liquid. The 
optimization procedure, required for the search of the domain of maximum probability, is carried out by the implementation of the level set method. 
Results for a couple of case studies are presented, to illustrate the structure of liquid water at ambient conditions and upon increasing pressure from the point of view of MPDs and to compare the information encoded in the solvation shells of sodium in water with, once again, that extracted from the MPDs.
\end{abstract}
\maketitle 

\section{Introduction}
While liquids have long-range disorder, they exhibit a local order that drives many of their physical properties. This short-range order is usually 
characterized through the Radial Distribution Functions (RDFs)~\cite{hansen}. However, the RDFs convey little or indirect information about the 
three-dimensional (3D) local structure of liquids. Spatial Distribution Functions (SDFs) have then been introduced for this 
purpose~\cite{svishchev_structure_1993, daub_structure_2009}. Since then, several indicators of the local order of liquids have been proposed, in 
particular for the important case of liquid water, including the statistics of Voronoi polyhedron~\cite{shih_voronoi_1994, jedlovszky_voronoi_1999, 
chakraborty_analysis_2011,  stirnemann_communication:_2012, idrissi_characterization_2013}, local structure index~\cite{errington_relationship_2001} 
or angular correlation~\cite{saitta, babiaczyk_hydration_2010}. Recently, local density fluctuations and the probability distribution of occupancy of 
a given volume have attracted a lot of attention, in the framework of the theory of hydrophobicity~\cite{chandler_interfaces_2005}. These have been 
used to characterize and locate patches of hydrophobicity or hydrophilicity at surfaces~\cite{acharya_mapping_2010, patel_quantifying_2011, 
rotenberg_molecular_2011} or around proteins~\cite{acharya_mapping_2010, patel_efficient_2014}. However, as such, they have not been used to 
characterize the microscopic structure of liquids.

In a very different context, that of electronic structure, related concepts have been used to locate electron pairs~\cite{cances_how_2004, 
francisco_electron_2007, pendas_pauling_2007}. In general, these methods are designed to identify regions in 3D space with particular chemical and 
physical meaning. The analysis is based on the information provided by the electronic probability density calculated from the many-body wave function, 
but offers a way to define and to visualize some relevant regions in real space.

The work presented in this paper is motivated by the interest for developing an alternative approach to study the local structure of liquid systems, 
combined with the clear similarities between the electronic probability density and the atomic many-body probability distribution of a liquid. We 
propose here a method based on the identification of regions of space where the probability of finding one and only one particle is maximum. We refer 
to these regions as Maximum Probability Domains (MPDs). The definition of this probability is given as the generalization of the 
concept~\cite{cances_how_2004, gallegos_maximal_2005, savin, causa_maximum_2011, lopes_jr_understanding_2012} used for the analysis of electronic 
probability densities in molecules and indeed can be generalized to the case of $n$ (and only $n$) particles. The information encoded in this 
quantity has a many-body nature, also when considering the one-particle occupancy probability. It has to be viewed as the probability of finding $n$ 
particles within a certain region of space with all other particles outside. Therefore, such probability is different from any reduced $n$-particle 
probability density, since the degrees of freedom of the $N-n$ remaining particles ($N$ being the total number of particles in the system) are not 
integrated out. In analogy to such quantities, however, the probability can be evaluated by using a standard sampling procedure with Molecular 
Dynamics (MD) or Monte Carlo (MC) trajectories.

We illustrate the method by applying it to describe the structure of pure liquid water and to define the solvation shells of sodium ions in a diluted 
water solution. We compare the results of our study with more standard approaches based on the RDFs and with various definitions of solvation shells. 
It will become apparent from the illustrations that the method offers an efficient complementary tool to the analysis of liquid structures, both 
qualitatively and quantitatively. The one-particle MPD defines the boundary of the 3D region available to a given particle, where it is not likely 
that another particle can penetrate. When a set of MPDs is identified around a central water molecule or sodium ion, we obtain a 3D-map of the 
locations of the surrounding molecules with a number of MPDs that is usually smaller, as we will see later, than the standard coordination number. The 
request of exclusive occupation of a MPD by only one particle is responsible for this feature, even though for rigid local structures, as 
the first solvation shell of water, there are clear similarities between the two pictures (MPDs and coordination numbers). 

The results for liquid water at different densities are mainly presented as a test case, in order to prove that the method is able to recover known results even if the problem is analyzed from a new perspective.  A more challenging application is represented by the case of sodium in water. Here, the one-particle MPDs located around the central ion are organized such that the spherical symmetry of the problem is maintained. However, the shape and dimension of the domains give a 3D resolution that is not accessible, or at least not directly, when employing the SDFs. We find relevant to stress here that, as will become clear from the applications, the major strength of the method developed in the paper is not to be searched in a new overall picture of the liquid achieved by determining the MPDs. The details are what make this method an interesting alternative or complement to more standard analysis techniques. Being able to pinpoint the location of the interstitial water molecules that are mainly affected by the increase of the density in liquid water or analyzing the shape of the regions occupied by water around a central molecule or ion could be used to predict properties such as the life-time of a H-bond or the kinetic of a reaction. These developments will not be discussed here, since our major goal is establishing the theoretical basis of the method and proving its efficiency with a few illustrations.

The paper is organized as follows. In Section~\ref{sec: probability}, we introduce the general definition of the one-particle occupancy probability. 
Then, the maximization of such probability is posed as a geometric optimization problem and solved within the framework of the Level Set Method (LSM) 
in Section~\ref{sec: level set method}. Section~\ref{sec: implementation} gives the details for the numerical implementation of the method and we 
present the two illustrations mentioned before in Section~\ref{sec: application}. Finally, in Section~\ref{sec: conclusion} we conclude and state our 
outlook for future developments.

\section{Definition of Maximum Probability Domains}\label{sec: probability}
\newcommand{\R}{\mathbf R}
\newcommand{\D}{\Delta}
\newcommand{\Prob}{P^{(1)}(\Delta)}
\newcommand{\dens}{\rho(\mathbf R^{N})}

The one-particle Maximum Probability Domain (MPD) is defined as the region in three-dimensional (3D) space where the probability of finding one, and 
only one, particle is maximum. In the present context, the term particle stands for an atom or the center of mass of a molecule or some other point 
of high symmetry in a molecule. The choice depends on the information that we want to extract from this analysis. For instance, in the study of 
liquid water, as presented in this paper, the \textsl{particles} will be the oxygen atoms in the molecules.

If a system is composed of $N+1$ identical particles with positions ${\R'}^{N+
1}=\R_{0}',\R_{1}',\ldots,\R_{N}'$ and we focus on the $N$ particles with positions $\R_i=\R_i'-\R_0'$ for $i=1,\ldots N$, we can define the 
one-particle occupancy probability of a domain $\D$, $\Prob$, as the probability of finding one, and only one, particle in the region of space $\D$ 
with all other particles located outside $\D$. This probability~\cite{savin}, written as
\begin{align}\label{eqn: def probability}
\Prob = \sum_{i=1}^N &\int_{\D_{c}} d\R_{1}\ldots d\R_{i-1}  
\int_{\D} d\R_{i}\int_{\D_{c}} d\R_{i+1}\ldots d\R_{N}\,\dens,
\end{align}
is what in mathematics is called a set function. Here, $\dens$ is the configurational probability density and, in the particular case of water and 
sodium diluted in water, will be explicitly defined in Appendix~\ref{app: probability}.

The domain $\D$ is given in a reference frame that is centered on the particle at $\R_0'$, which is kept out of statistics (in the sense that in 
Eq.~(\ref{eqn: def probability}) we do not integrate over its positions). $\D_c$ is the symbol used to indicate the complementary volume of $\D$. The 
sum in Eq.~(\ref{eqn: def probability}) allows us to define the probability that is independent of the identity of the particle occupying the domain 
$\D$. Moreover, the notation in Eq.~(\ref{eqn: def probability}) indicates that all integrals involving the variables $\R_{1}\ldots \R_{i-1}$ and 
$\R_{i+1}\ldots \R_{N}$  are performed over $\D_c$. This definition of the probability can be easily extended to $n$ particles in $\D$ and $N-n$ 
particles in $\D_c$. Eq.~(\ref{eqn: def probability}) satisfies the normalization condition
\begin{align}\label{eqn: normalisation of the probability}
\sum_{n=0}^N P^{(n)}(\D) = 1,
\end{align}
which can be easily obtained for an ideal gas (in the absence of interactions among particles) occupying a volume $V$, as the configurational 
probability density is explicitly known, i.e. $\dens=V^{-N}$, thus allowing for the analytic calculation of Eq.~(\ref{eqn: def probability}). This is 
done by inserting the expression of the one-particle occupancy probability $P^{(n)}(\D)=\binom{N}{n}v^n(V-v)^{N-n}/V^N$ in Eq.~(\ref{eqn: 
normalisation of the probability}), where $v$ is the volume occupied by $\D$ and $V-v$ that occupied by $\D_c$.

To extend the integration domain in Eq.~(\ref{eqn: def probability}) to the whole configuration space, we introduce the \textsl{characteristic 
function} $\Upsilon_{\D}(\R)$, defined as
\begin{equation}
\Upsilon_{\D}(\R) =\left\lbrace
\begin{array}{ll}
1 & \mbox{if}\,\,\R\in \D\\
0 & \mbox{otherwise}
\end{array}
\right.,\quad \Upsilon_{\D_{c}}(\R) = 1-\Upsilon_{\D}(\R).
\end{equation}
The probability $\Prob$ is then
\begin{equation}\label{eqn: prob with characteristic function}
\Prob = 
\sum_{i=1}^N \int d\R^{N} \Upsilon_{\D}(\R_i)\prod_{j\neq i}^{N} \left(1-\Upsilon_{\D}(\R_{j})\right)\dens
\end{equation}
which allows us to identify the microscopic observable
\begin{equation}
\Gamma_{\D}^{(i)} (\R^{N})=\Upsilon_{\D}(\R_{i})\prod_{j\neq i}^N \left(1-\Upsilon_{\D}(\R_{j})\right).
\end{equation}
The equilibrium average of $\Gamma_{\D}^{(i)} (\R^{N})$, evaluated according to the probability density $\dens$, leads to the definition of $\Prob$
\begin{equation}\label{eqn: prob as an equilibrium average}
\Prob = \left\langle \sum_{i=1}^{N}\Gamma_{\D}^{(i)} \left(\R^{N}\right) \right\rangle.
\end{equation}
Eq.~(\ref{eqn: prob as an equilibrium average}) can be evaluated by sampling the microscopic observable $\Gamma_{\D}^{(i)} (\R^{N})$ along a 
Molecular Dynamics (MD) or Monte Carlo (MC) trajectory. From the algorithmic point of view, $\Prob$ can be estimated by computing
\begin{equation}\label{eqn: numerical sample of prob}
 \nu^{(1)}(\D)=\frac{1}{N_{conf}} \sum_{\kappa=1}^{N_{conf}}\delta_{1,n_\D(\kappa)}\simeq\Prob,  
\end{equation}
the frequency of the event ``one, and only one, particle inside $\D$''. Here, $N_{conf}$ is the total number of configurations sampled along the 
trajectory and $\kappa$ labels the selected configuration. The term in the sum is equal to one only if the number of particles $n_\D(\kappa)$ inside 
the volume $\D$, for the configuration $\kappa$, is equal to 1.

The construction of the MPDs is carried out by initially choosing some centers around the central particle (labeled by 0) and some volumes enclosing 
these centers. A set of MPDs will be then identified as
\begin{equation}\label{eqn: mpd}
\D^{*}_i=\arg\max_{\D_i} P^{(1)}(\D_i), \quad\mbox{with} \,\,i=1,2,\ldots.
\end{equation}
The centers around the 0th particle can be chosen quite arbitrarily, however we find useful to locate them in correspondence of the local maxima of 
the Spatial Distribution Function (SDF) or of the Radial Distribution Function (RDF). It is worth underlining that $\D^*_i$ is not an absolute 
maximum~\footnote{The method does not 
determine the MPD whose one-particle occupancy probability, $\Prob$, is the absolute largest, but finds several MPDs 
associated to different values of $\Prob$ separated by regions where this probability has (local) minima.} over all space of volumes, but it is the 
one found by a local 
search (the closest to the initial choice).

The definition of the set of MPDs $\D_{1}^{*},\D_{2}^{*},\ldots$ can be used for the characterization of the structure of atomic and molecular 
liquids as an alternative or a complement to standard tools such as distribution functions, solvation shells and coordination numbers. The 
determination of the optimized domains $\D^{*}_i$'s is posed as a geometric optimization problem, since the probability $P^{(1)}(\D_i)$ has to be 
maximized with respect to variations of $\D_i$. In the following section we present~\cite{shape-top_der, optimal_shape, sensitivity_analysis, 
inside-out, propagating_fronts} the Level Set Method (LSM) to be applied in the search of the MPDs, in particular in a liquid.

\section{Shape derivative in the level set method}\label{sec: level set method}
A central concept in the procedure referred to as LSM is that of shape derivative. The function $\Prob$ belongs to a specific class of 
what is normally called a set function and the variations of a set function with respect to changes of the set define its shape derivative. The aim 
of the numerical procedure developed and tested in this paper is to determine how $\Prob$ changes by varying $\D$, such that the value of the 
probability itself is maximized. Appendix~\ref{app: lsm} is devoted to a detailed discussion on the shape derivative and on its expression in the 
context of the MPDs. Here, we will focus on illustrating the essence of such concept.

The set function defined in Eq.~(\ref{eqn: def probability}) depends on the integration volume, $\D$. Therefore, when $\D$ changes, also the value of 
the integral performed over $\D$ changes: such (infinitesimail) variations of $\Prob$ will be indicated with the symbol $\shape\Prob$, the shape derivative. In the 
particular application discussed here, we request that this variation of the integration domain follows a well-defined law: The value of the integral 
has to increase as $\D$ is varied. This choice of variation allows us to define a \textsl{deformation law}, namely a fictitious equation of 
motion such as
\begin{equation}\label{eqn: deformation law in the text}
\D\rightarrow \D_{\tau} = \left\lbrace \R_{\tau}=\R+\V d\tau \,|\,\R\in \partial\D \right\rbrace,
\end{equation}
with $\partial\D$ the border of $\D$. The evolution, or deformation, of the domain, where $d\tau$ is a fictitious time-step, is completely defined by 
the fictitious velocity field $\V$. The explicitly expression of the velocity field is determined in Appendix~\ref{app: lsm}, by requiring that 
$\shape\Prob\geq0$. The maximization condition on the shape derivative enables us to optimize the domain $\D$ towards reaching the maximum of the set 
function $\Prob$.

The numerical procedure developed to evolve the domain $\D$ towards reaching the maximum value of $\Prob$ is given in Appendix~\ref{app: lsm}, while in Section~\ref{sec: implementation} we implement the algorithm based on the LSM to compute the MPDs in a liquid system.

\section{Numerical implementation}\label{sec: implementation}
\subsection{Choice of the initial volumes and Bader partitioning}
To apply the LSM we need to choose a starting set. In the two cases of pure water or sodium diluted in water we locate the initial domains $\D_i$ in 
correspondence to the local maxima of the oxygen-oxygen SDF or of the sodium-oxygen RDF, respectively. This difference is related to the fact that 
the SDF of a sodium ion is spherically symmetric.

The two-particle SDF~\cite{gray-gubbins} $\rho^{(2)}(\R)$, employed here and calculated around a central (0th) water molecule, is proportional to the probability of 
finding an oxygen atom in $\R$ irrespective of the position of all other oxygen atoms~\cite{hansen}, given a certain configuration $\lbrace 
\R_0,\R_0^{\mathrm H_1},\R_0^{\mathrm H_2}\rbrace$ of the central molecule
\begin{align}
\rho^{(2)}(\R)\equiv\rho^{(2)}\left(\R\left.\right|\R_0,\R_0^{\mathrm H_1},\R_0^{\mathrm H_2}\right)
=N(N-1)\int d\mathbf R^{N-1}\rho\left(\R,\R^{N-1}\left.\right|\R_0,\R_0^{\mathrm H_1},\R_0^{\mathrm H_2}\right).\label{eqn: OO SDF}
\end{align}
The probability density under the integral sign is a conditional probability density,
\begin{align}
\rho\left(\R,\R^{N-1}\left.\right|\R_0,\R_0^{\mathrm H_1},\R_0^{\mathrm H_2}\right)
=\int d{\R^{\mathrm H_1}}^N \, d{\R^{\mathrm H_2}}^N\,\frac{\rho^{can}\left(\R_0,\R,\R^{N-1},\R^{\mathrm H_1}_0,\R^{\mathrm H_2}_0,{\R^{\mathrm 
H_1}}^N_1,{\R^{\mathrm H_2}}^N\right)}{P_m\left(\R_0,\R^{\mathrm H_1}_0,\R^{\mathrm H_2}_0\right)},
\end{align}
exactly as in Eq.~(\ref{eqn: conditional probability}), with $\rho^{can}$ the standard configurational canonical density and $P_m\left(\R_0,\R^{\mathrm H_1}_0,\R^{\mathrm H_2}_0\right)$ the marginal probability of finding the central molecule in the 
configuration $\lbrace\R_0,\R^{\mathrm H_1}_0,\R^{\mathrm H_2}_0\rbrace$, given by Eq.~(\ref{eqn: 
marginal probability}). Note that $\rho^{(2)}(\R)$ is now a function of the three cartesian coordinates $\R$ and can therefore be visualized in 3D 
space, as we will show below.

By employing Bader analysis~\cite{bader}, we locate maxima, and domains around them, of the two-particle SDF in Eq.~(\ref{eqn: OO SDF}). In doing 
that, we apply an analysis method designed for the study of electronic probability densities to the study of a liquid. Bader analysis partitions the 
space in regions assigned to the maxima of the density. From these regions starts the search of the MPDs which is not a partition of the space and can 
result in overlap between domains or partial filling of the spaces.

Exactly as it is done for Bader partitioning the electronic density, the Bader procedure applied to the two-particle SDF is based on the assumption that the SDF can be written as the sum 
of $L$ independent contributions,
\begin{align}
\rho^{(2)}(\R) = \sum_{\alpha=1}^L \rho^{(2)}_\alpha(\R),
\end{align}
where $L$ is the number of total domains to be determined by the procedure and is equal to the number local maxima (modes) of the density. In writing this relation, we make the further hypothesis that $\rho^{(2)}_\alpha(\R)\neq 0$ only if $\R\in \D_\alpha$, with $\D_\alpha$ a volume in 3D space. The main objective here is to determine 
the $\D_\alpha$'s, referred to as Bader domains, and to use them as initial conditions for the optimization procedure based on the LSM. Since we are 
working in 3D space, Bader domains are enclosed by the 2D surfaces where the density gradient has zero flux. In actual calculations, to fully 
identify the Bader domains, given an initial $\R'$, we compute the vector $\R=\R'-\R_0$, where $\R_0$ is the origin of our reference system, and we 
evaluate the gradient of $\rho^{(2)}(\R)$ at that point. Then we construct a steepest-ascent path in 3D space whose tangent in every point is the 
direction of the gradient. When the maximum is reached, all points of that path are associated to a Bader volume. Another point $\R'$ is then chosen 
and the procedure is iterated until all space points have been assigned to a specific maximum. Indeed, starting from different points, the same 
maximum can be reached thus the Bader volume has to be updated.

The main reason for choosing Bader partitioning as initial condition for our optimization procedure is efficiency. The density 
landscape surrounding a water molecule is very complex and not homogeneous. Therefore, starting with initial domains that already contain information 
about the local structure of the liquid does help in efficiently determining the MPDs. Employing Bader analysis is, however, not possible in 
general, but it depends on the topology of the two-particle SDF. In fact, this approach fails in the case of the spherically symmetric 
sodium-oxygen SDF and we have to proceed differently. We take as initial sets few spheres of different radii, generally smaller than 
2.5~\AA, whose centers are approximately located at a distance from sodium corresponding to the two main maxima of the sodium-oxygen RDF, i.e. at 
$\sim 2.5$~\AA~and $\sim 4.5$~\AA~from the sodium ion (see for instance Fig.~\ref{fig: rdf Na-water}).

\subsection{Algorithm}
According to the steps described in Section~\ref{sec: level set method}, we
\begin{enumerate}
\item define the initial domain $\D_i^{0}$;
\item construct a level set function. Our choice~\cite{propagating_fronts} is
\begin{equation}
\phi(\R,0)=\left\lbrace
\begin{array}{ll}
-1+\exp[-d(\R)/\sigma] & \mbox{if} \,\,\R\in\D \\
0 & \mbox{if} \,\,\R\in\partial\D \\
1-\exp[-d(\R)/\sigma] & \mbox{if} \,\,\R\not\in\D
\end{array}
\right.
\end{equation}
where $d(\R)$ is the shortest distance of the point $\R$ from $\partial \D$ and $\sigma$ is a parameter to be chosen;
\item calculate $\Prob$ from Eq.~(\ref{eqn: numerical sample of prob});
\item calculate $f_\D(\R)$ by sampling the microscopic observable in Eq.~(\ref{eqn: 
f_D});
\item evolve the level set function according to Eq.~(\ref{eqn: evolution lsf}); 
\item go back to point 3. and iterate until the variation of the probability $\Prob$ is smaller than a certain threshold $\delta$, i.e. 
$|P^{(1)}(\D_\tau)-\Prob|\leq\delta$.
\end{enumerate}
The fictitious time-step $d\tau$ to be used in our calculations has resulted to be 10.0, with a value for $\sigma$ of 0.1~\AA~while the threshold to 
monitor the convergence of the algorithm has been chosen to be $\delta=10^{-7}$. Tests have been performed with different values of the 
convergence parameter and no significant changes have been observed in the final results if $\delta$ is chosen to be $10^{-5}$ at most. A good grid 
spacing for the 3D visualization of the MPDs has resulted to be 0.2~\AA.

\section{Case studies}\label{sec: application}
In the first solvation shell of pure water at suitable conditions defined later on, Bader domains are shown in Fig.~\ref{fig: bader basins, 1st 
shell}, along with $\rho^{(2)}(\R)$.
\begin{figure}[h!]
\begin{center}
\includegraphics*[width=.45\textwidth]{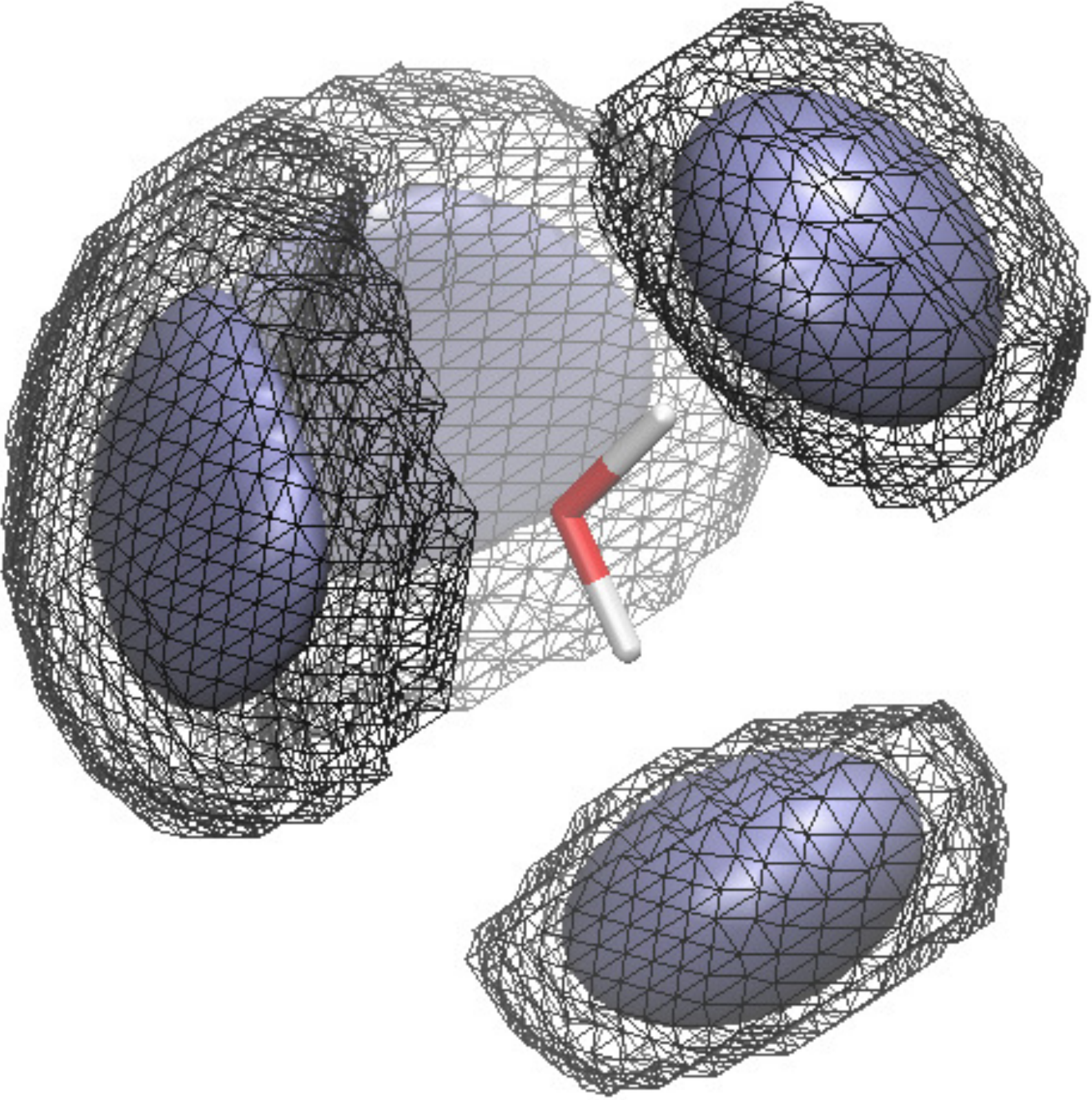}
\caption{The filled areas represent the two-particle SDF computed for water at density $\rho_0=1.0$ g/cm$^3$ (rendering of contour surface 0.17 of $\rho^{(2)}(\R)/\rho_0$). The grids enclose the Bader 
domains, namely the volumes which are associated to the four highest maxima of the two-particle SDF obtained by the application of Bader analysis.}
\label{fig: bader basins, 1st shell}
\end{center}
\end{figure}
If we restrict ourselves to the four highest maxima of $\rho^{(2)}(\R)$ computed for water, Bader analysis identifies the domains $\left\lbrace\D_i 
\right\rbrace_{i=1,4}$ that are represented by the volumes enclosed by the grids in Fig.~\ref{fig: bader basins, 1st shell}. The filled areas are 
shown for reference and represent the function $\rho^{(2)}(\R)$. For each domain, the probability of finding one, and only one, particle 
$P^{(1)}(\D_i)$ is calculated from Eq.~(\ref{eqn: numerical sample of prob}) and then maximized, by following the procedure illustrated above. The 
final domain is the MPD available to each particle surrounding the central water molecule.

\begin{figure}
\begin{center}
\includegraphics*[width=.45\textwidth]{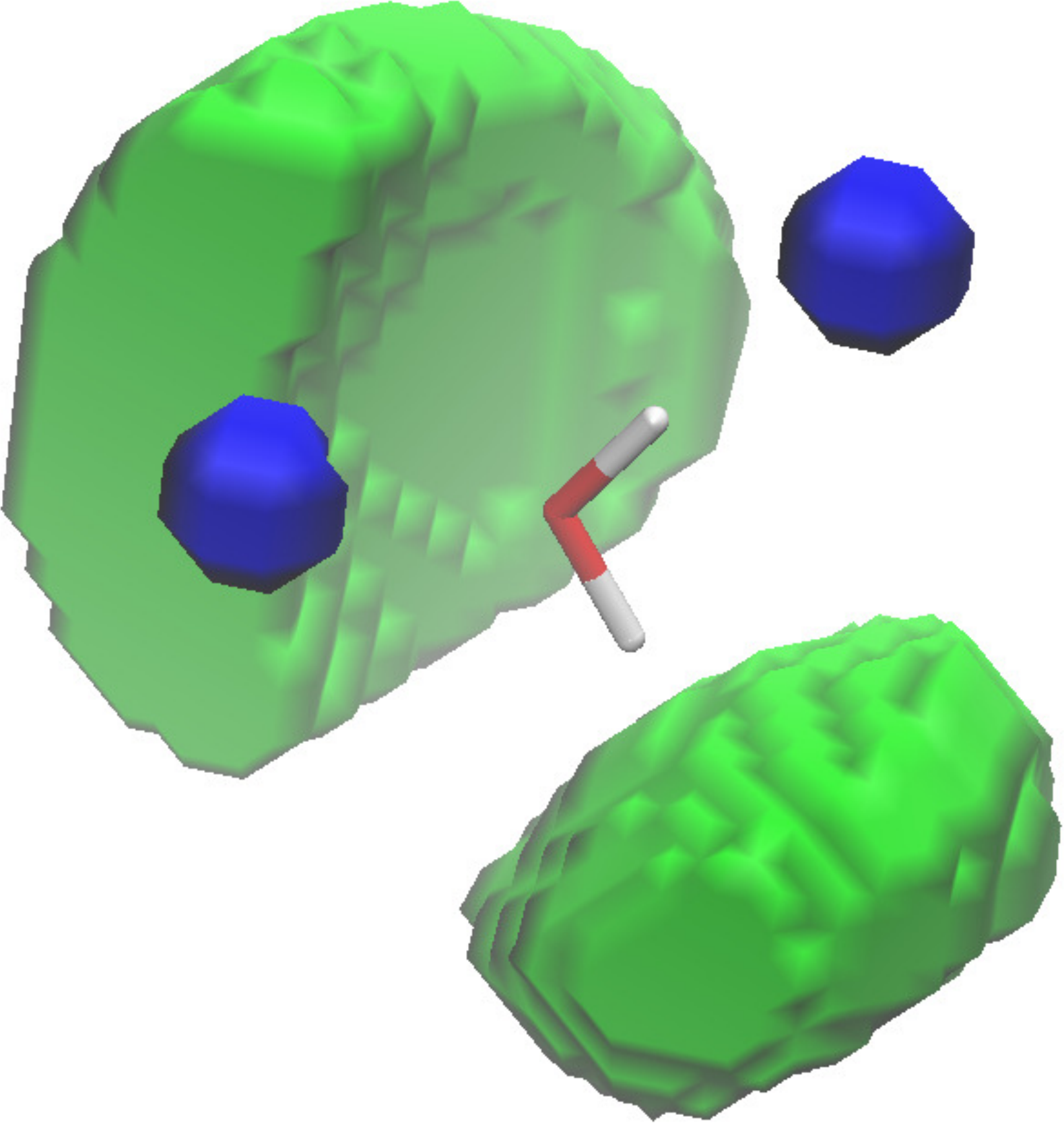}
\includegraphics*[width=.45\textwidth]{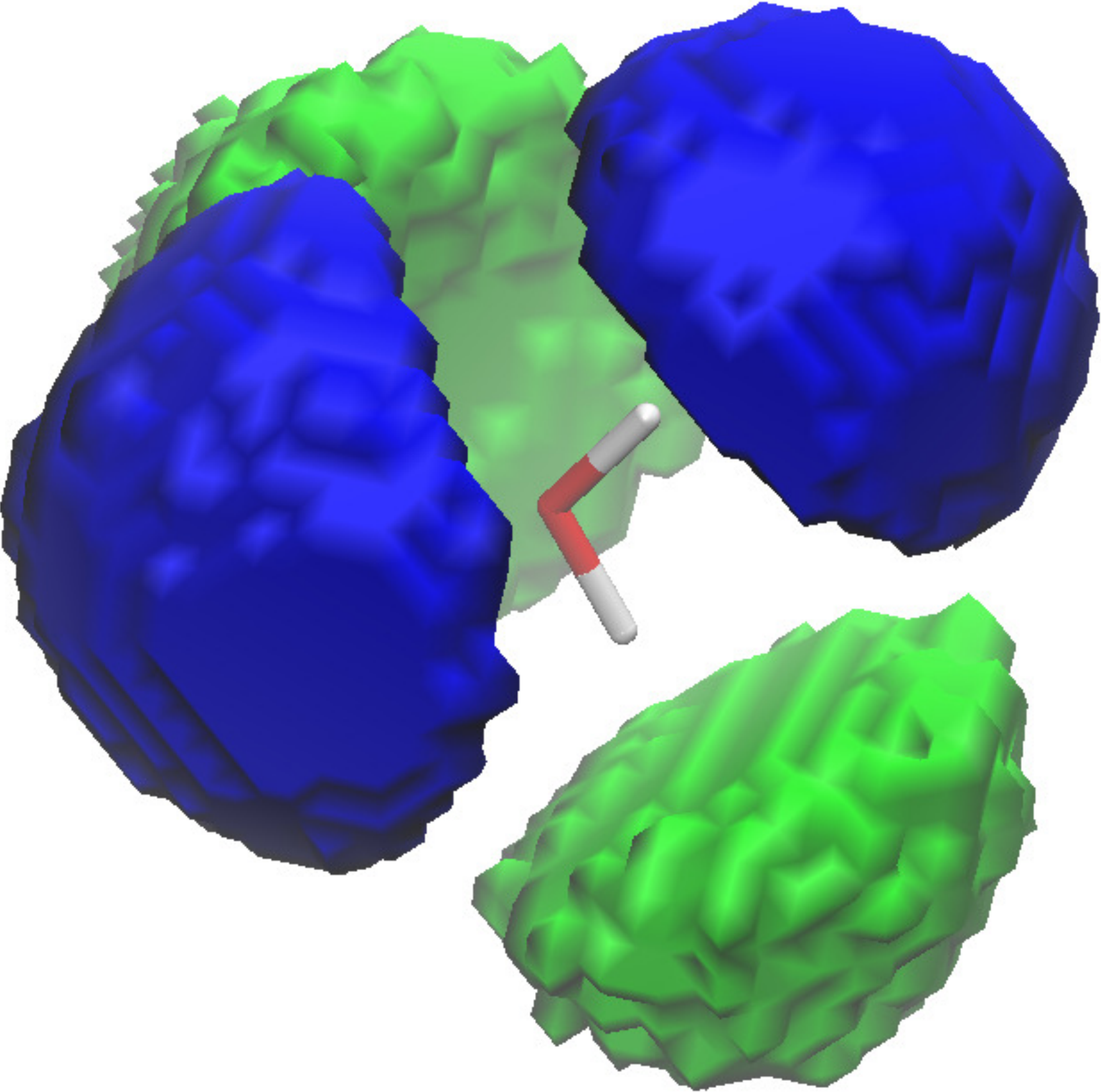}
\caption{Left: different initial conditions for two sets of equivalent domains around a water molecule. Two domains are located in the region of the 
first solvation shell corresponding to the two oxygen atoms accepting the hydrogen bond; two domains appear in the region in the first solvation 
shell 
occupied by the two water molecules donating the hydrogen bonds. The green areas are Bader domains, whereas the blue areas are spheres with centers 
located at the positions of the maxima of the two-particle SDF. Right: the domains located at equivalent positions are identical after optimization.}
\label{fig: initial conditions}
\end{center}
\end{figure}

The applications of the LSM are illustrated below. Before doing that, we find useful to stress two points. \\
(i) The results of our analysis do not depend on the initial set chosen. To that end, let us look at the result of the optimization performed using 
two different initial conditions for maxima of equivalent physical meaning. The first type of initial condition is a Bader domain (green areas in 
Fig.~\ref{fig: initial conditions}, left). The second, for different but equivalent maxima, is a sphere, approximately centered at the positions of 
the 
maxima of the SDF (blue areas in Fig.~\ref{fig: initial conditions}, left). Fig.~\ref{fig: initial conditions} (right) shows that the shape of the 
MPDs is independent of the initial conditions.
\\
(ii) When Bader analysis is applied in order to identify the initial domains, the space around the central particle (water molecule in our case) is 
partitioned as far as the two-particle SDF shows well defined local maxima. When the liquid approaches a random distribution Bader analysis becomes 
inefficient. In the case of water at density $\rho_0=1.0$ g/cm$^3$ and $\rho=1.23$ g/cm$^3$ presented in two following sections, we are able to 
identify and analyze domains up to within the third solvation shell ($6 - 8$~\AA).

An additional point that is worth discussing here is the statistics. In order to increase the accuracy of our calculations in the case 
of pure water (Sections~\ref{sec: water 1.0} and~\ref{sec: water 1.23}), we observe that all molecules are identical. Therefore, each molecule can be used as the central one, with respect to which the 
one-particle occupancy probability is calculated. This procedure is employed to overcome problems related to a short trajectory simulation. This 
same operation is, however, not possible in the case of sodium (Section~\ref{sec: sodium in water}), because only one ion is present in the simulation box. In this case, since the 
problem has spherical symmetry, we increase the statistics by averaging over different orientations of the reference system centered on the ion.

\subsection{Liquid water at {\boldmath$\rho_0=1.0$ g/cm$^3$}}\label{sec: water 1.0}
The MD trajectory of 150~ps for liquid water at room temperature is generated by employing the TIP4P~\cite{tip4p} model. The system is composed of 
4096 molecules in a cubic box with side length 49.7~\AA. Periodic boundary conditions are used throughout.

The 3D two-particle SDF has been calculated from the MD trajectory and, from the application of Bader analysis, the space around a central water 
molecule is partitioned in 15 regions within a distance of about 8~\AA~from the central oxygen atom. The resulting 15 MPDs give information on the 
structure of water up to within the third solvation shell, as we will now show.

Results will be presented by first looking at the MPDs in comparison to Bader domains. A 3D-map of the distribution of oxygen atoms 
surrounding the central water molecule is obtained. Each MPD is, on average, occupied by a single oxygen, thus such 3D-map has a very clear 
physical interpretation, pinpointing the locations of the oxygen atoms around a given water molecule. After this preliminary, more qualitative, 
presentation of the results, a few analysis tool will be introduced: we will classify the MPDs according to the corresponding value of the 
one-particle occupancy probability, according to their distance from the central oxygen atom and according to their volumes. These properties will be 
used to rationalize the differences between water at $\rho_0=1.0$ g/cm$^3$ and at higher density, $\rho=1.23$ g/cm$^3$, presented in the following 
section. Finally, the results shown here will be compared to more standard analysis tools, like RDFs, angular distributions and hydrogen bond 
structure.

\subsubsection{Qualitative analysis of the MPDs in water}
\begin{figure}[h!]
\begin{center}
\includegraphics*[width=.4\textwidth]{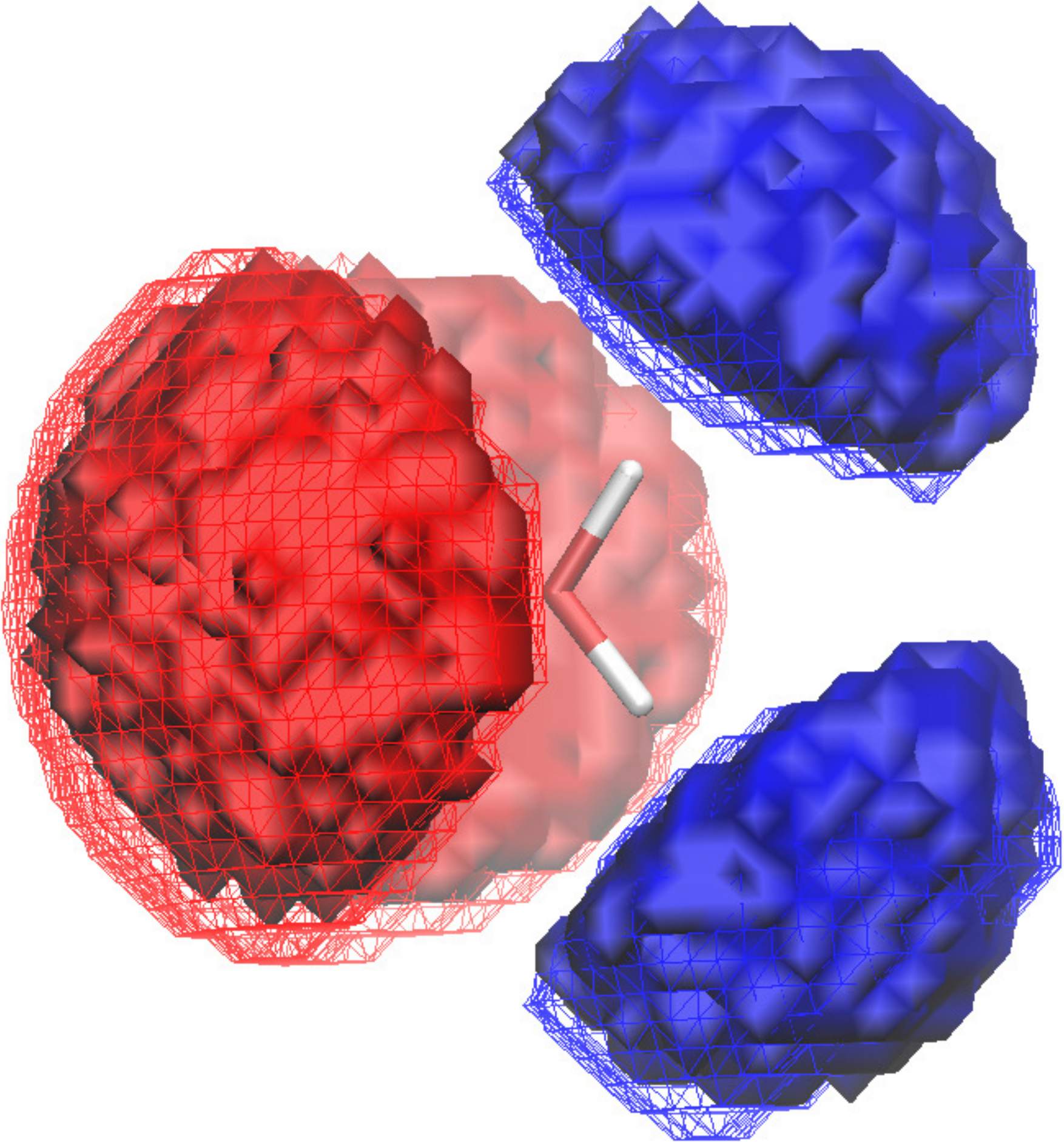}
\caption{Case of water at $\rho_0=1.0$ g/cm$^3$. MPDs (filled areas) labeled 1, 2 (blue) and 3, 4 (red) around the central water molecule. The initial Bader domains are also 
shown as grids.}
\label{fig: basins optimisation 1 to 4}
\end{center}
\end{figure}

\begin{figure}[h!]
\begin{center}
\includegraphics*[width=.45\textwidth]{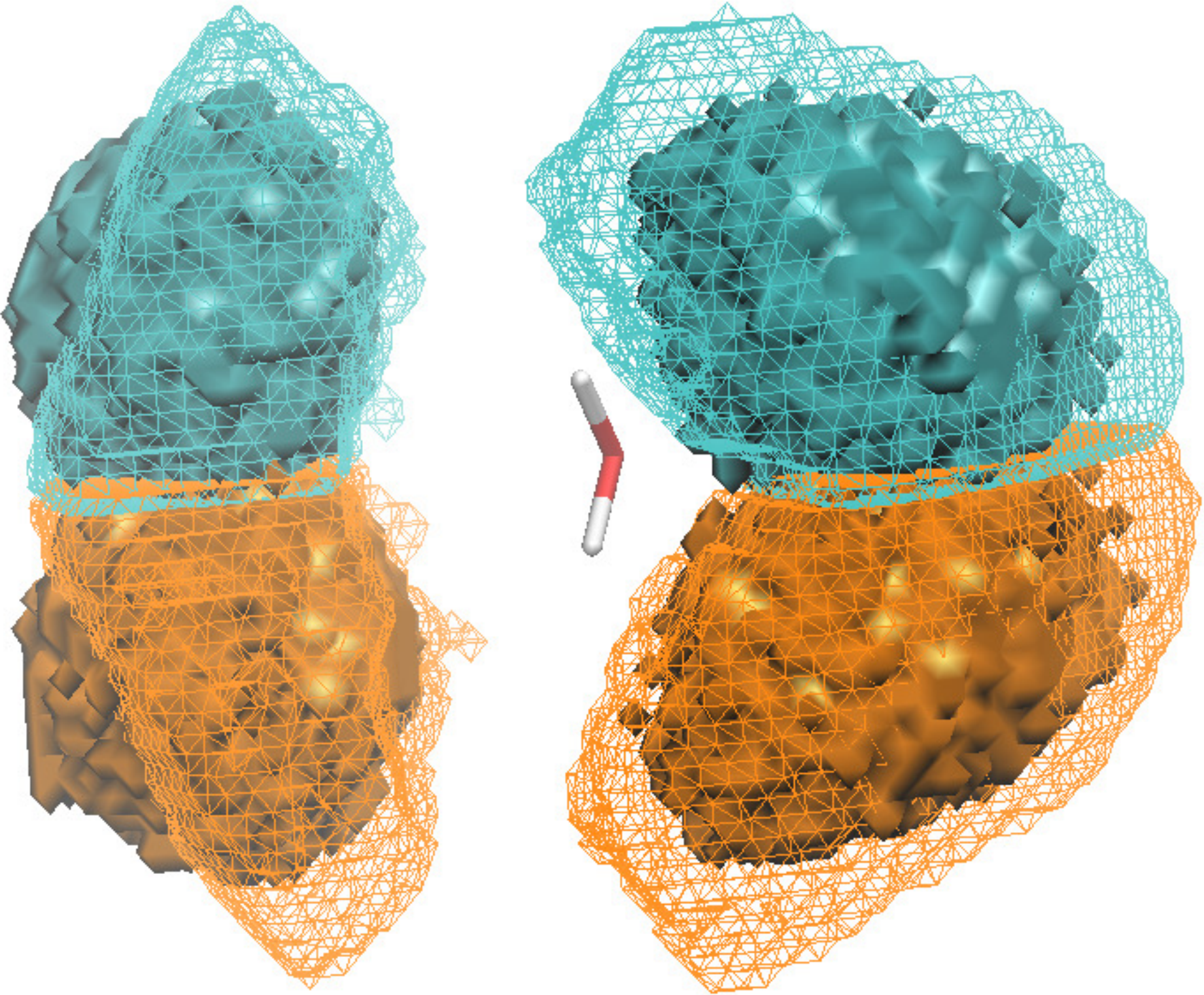}
\includegraphics*[width=.38\textwidth]{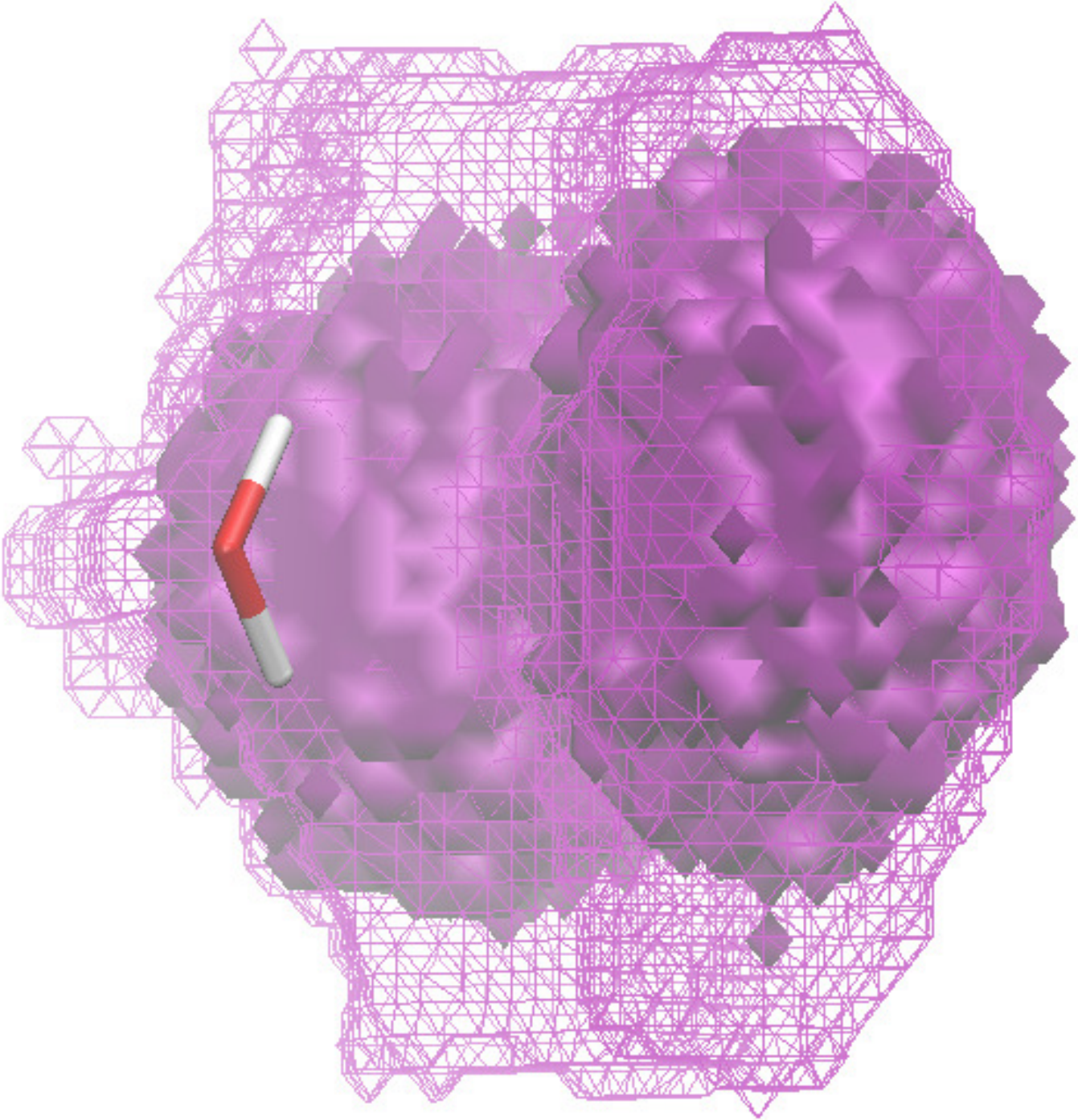}
\caption{Case of water at $\rho_0=1.0$ g/cm$^3$. Left: MPDs labeled 5, 6 (cyan) and 7, 8 (orange). Right: MPDs labeled 9, 10 (violet). The MPDs are compared to the initial 
Bader domains, represented as grids.}
\label{fig: basins optimisation 5 to 10}
\end{center}
\end{figure}

\begin{figure}[h!]
\begin{center}
\includegraphics*[width=.35\textwidth]{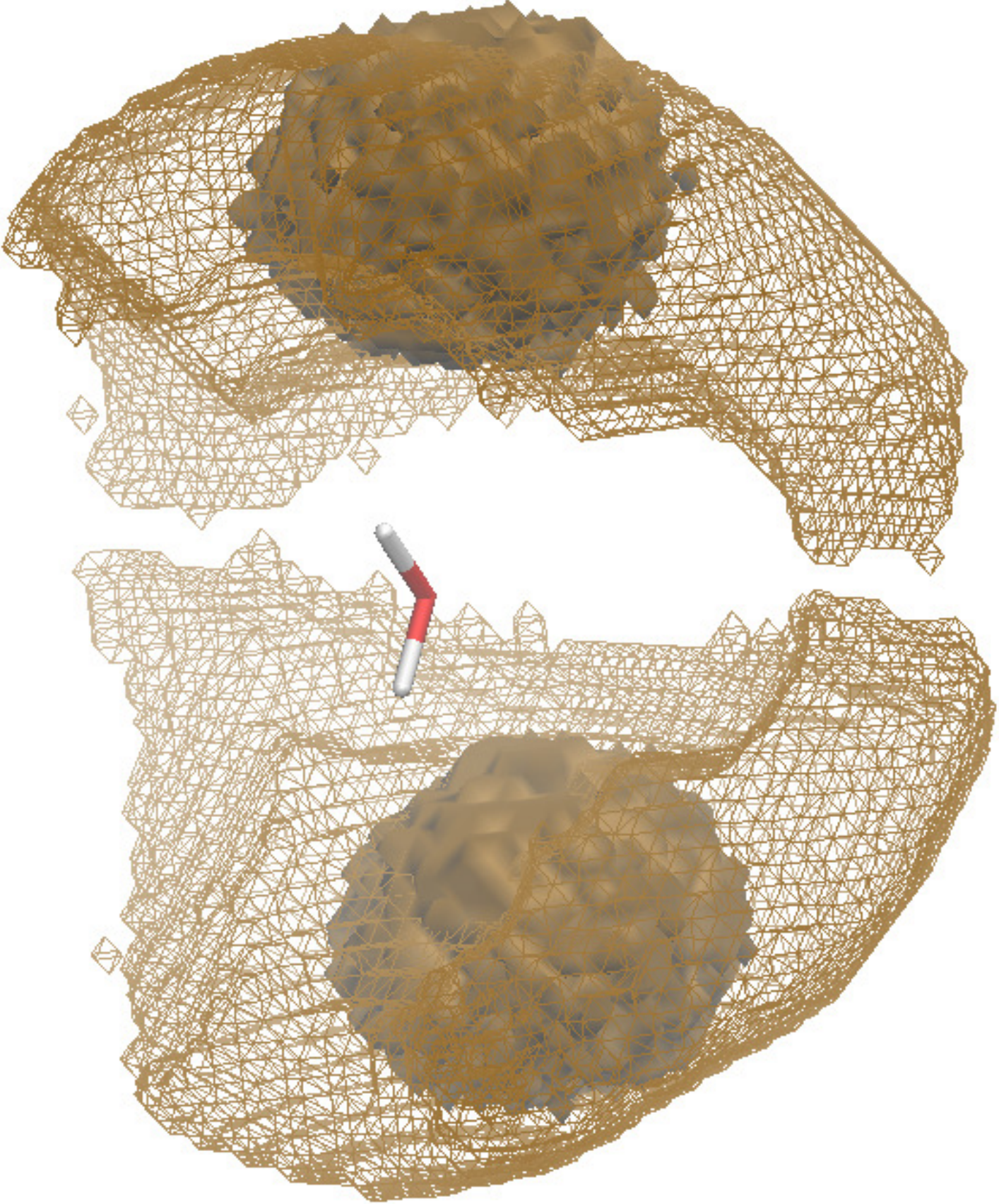}
\includegraphics*[width=.35\textwidth]{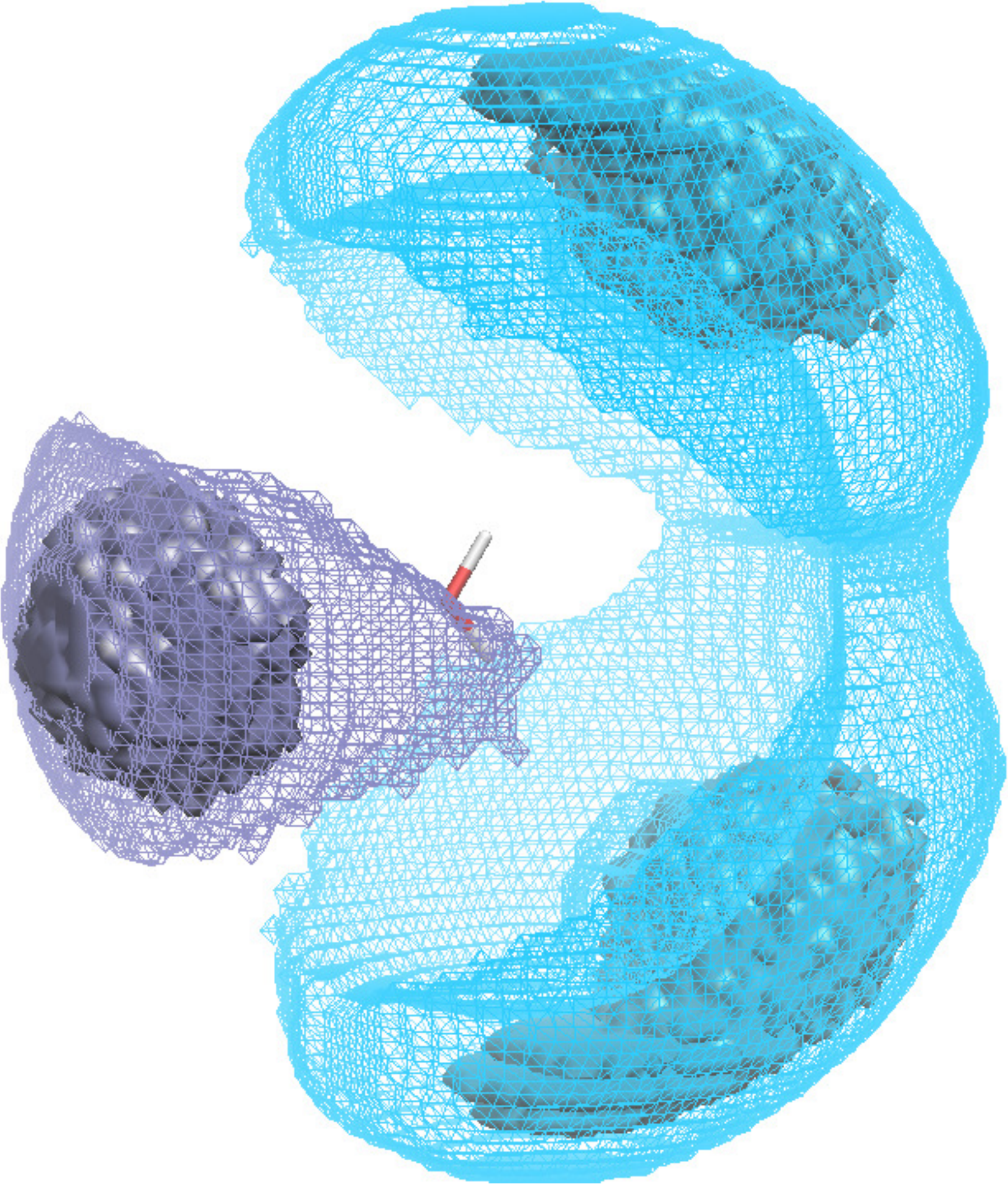}
\caption{Case of water at $\rho_0=1.0$ g/cm$^3$. Left: MPDs labeled 11, 12 (ochre areas). Right: MPDs labeled 13 (light-purple) and 14, 15 (turquoise). As in the previous 
figures, the initial Bader domains are shown for reference. We underline here that both pairs of domains 11, 12 and 14, 15 (Bader and MPDs) are 
completely symmetric with respect to the plane of the molecule and to its perpendicular plane.}
\label{fig: basins optimisation 11 to 15}
\end{center}
\end{figure}

Figs.~\ref{fig: basins optimisation 1 to 4},~\ref{fig: basins optimisation 5 to 10} and~\ref{fig: basins optimisation 11 to 15} compare the results 
from Bader partitioning the space around a water molecule from the MPD analysis. In all figures, Bader domains are indicated as grids, while the MPDs 
are the filled regions.
The four MPDs in Fig.~\ref{fig: basins optimisation 1 to 4} clearly identify the regions occupied by the four water molecules hydrogen-bonded 
(H-bonded) to the central molecule, arranged according to the typical tetrahedral structure of the first solvation shell of water.

Fig.~\ref{fig: basins optimisation 5 to 10} shows the domains 5 to 10: domains 5 to 8 (left) are arranged on both sides of the plane defined 
by the central molecule in front of the hydrogen atoms and are symmetric with respect to this plane; domains 9 and 10 (right) are located in front of 
the hydrogen atoms of the central molecule and, after the optimization, are similar to domains 5 to 8.

Domains 11 and 12 are shown in Fig.~\ref{fig: basins optimisation 11 to 15} (left) and, as will be proven below, they only partially occupy the first 
solvation shell, but they are mainly found in the second solvation shell, similarly to the domains 5 to 10. The MPDs 5 to 12 are \textsl{interstitial} 
domains, as they are located in correspondence of the empty spaces between the MPDs 1 to 4. 
The MPD labeled 13 (light-purple) in Fig.~\ref{fig: basins optimisation 11 to 15} (right) is also located in the second solvation shell, while 
further from the central molecule, the MPDs labeled 14, 15 (turquoise) are found as shown in Fig.~\ref{fig: basins optimisation 11 to 15} (right).

\subsubsection{Some tools for a quantitative analysis}
The MPDs around the central water molecule can be classified according to the value of the probability $\Prob$ in each domain, as shown in 
table~\ref{tab: distances}.
\begin{table}
\begin{tabular}{cccccc}
\hline
\hline
Label & $\left\langle N\right\rangle_{\textrm{before}}$ & $\left\langle N\right\rangle_{\textrm{after}}$ & Volume (\AA$^{3}$) & Distance (\AA) 
& $P^{(1)}(\D^*)$\\
\hline
1, 2 & 1.2 & 1.1 & 14.9 & 2.8 & 0.9,0.91 \\
3, 4 & 1.6 & 1.0 & 15.2 & 2.6 & 0.78 \\
5 - 8 & 2.0 & 1.2 & 15.8 & 4.2 & 0.63 \\
9, 10 & 1.8 & 1.1 & 15.9 & 4.3 & 0.63 \\
11, 12 & 5.9 & 1.1 & 15.8 & 4.4 & 0.63 \\
13 & 3.0 & \textit{1.8}$^{(a)}$ & 17.2 & 4.5 & \textit{0.37}$^{(a)}$ \\
14, 15 & 4.0 & \textit{0.6}$^{(a)}$ & 21.1 & 6.5 & \textit{0.61}$^{(a)}$\\
\hline
\hline
\end{tabular}
\caption{\label{tab: distances}List of the MPDs (first column), average number of particles in each volume before (Bader analysis, second column) or 
after the optimization (third column), volumes of the MPDs (fourth column), distances of the centers of mass of the MPDs from the central oxygen 
atom (fifth column) and the one-particle occupancy probability (sixth column). In the same line of the first column we have put equivalent domains. 
$^{(a)}$The values of $\left\langle N\right\rangle_{\textrm{after}}$ and $P^{(1)}(\D^*)$ for the volumes 13 and 14, 15 are in italic to indicate that 
the optimization procedure seems to be in these cases less efficient than for the other domains: in the first case, the optimization appears not able 
to split the domains in two parts; in the second case, the density of the liquid is probably too flat, thus preventing from an efficient analysis.}
\end{table}
The domains labeled 1, 2 are associated to a higher probability than the domains 3, 4. This difference can be explained as follows. Domains 1, 2 
enclose the oxygen atoms of the two water molecules H-bonded to the hydrogen atoms of the central water molecule, namely HO$\cdots$H$_{central}$. 
Domains 3, 4 enclose the oxygen atoms of the molecules that are H-bonded to the central one as OH$\cdots$O$_{central}$. Therefore, in the latter case, 
the oxygen atoms are not directly bonded to the central molecule, thus resulting more mobile.

The probability of finding only one water molecule inside domains 5 to 12 is further reduced compared to the previous case. It follows that a 
high probability is an indication of a strong bond, like a H-bond, while a lower probability suggests the absence of a strong interaction, as in the 
case of the oxygen atoms in the interstitial domains.

Table~\ref{tab: distances} shows also the volumes associated to each MPDs and their distance from the oxygen atom of the central water molecule, 
expressed in terms of the distance of their centers of mass. In particular, we notice the clear identification of three groups of domains, whose 
distances from the central molecule are below 3~\AA, between 4 and 5~\AA, above 6~\AA. We will see below how this structure changes upon increasing 
the density. We also observe the increase of the volume of the domain itself as its distance from the central oxygen increases. The correlation to 
the central water molecule is reduced, then the surrounding molecules are more mobile and occupy larger volumes. Observations about the effect of the 
interactions on the volume of the MPDs are reported in Appendix~\ref{app: ideal gas}.

\subsubsection{Comparison with standard analysis tools}
Since the domains 1 to 4 enclose the oxygen atoms of the water molecules that are H-bonded to the central molecule, we can define an estimate of the 
number  $n_{\mathrm{HB}}$ of hydrogen bonds (HBs) formed by a water molecule. Indeed the domains indicate only the positions of the oxygen atoms, and 
therefore, domains 3 and 4 cannot, strictly speaking, be used to estimate the presence of HBs. This is because in those regions the hydrogen atoms 
(not the oxygens) form HBs with the central water molecule. However, the H-bonded hydrogens are close to the oxygens occupying the domains 3 and 4. 
If we take a high one-particle occupancy probability as an indication of the presence of a HB, then its value can qualitatively measure the fraction 
of HBs present. Taking that the four closest domains do not superpose (a hypothesis almost always satisfied, but certainly an 
approximation~\cite{luzar-chandler}), we can sum up their probabilities to have an indication of how many HBs are formed around a water molecule:
\begin{equation}\label{eqn: number of HBs}
n_{\mathrm{HB}}=\sum_{i=1}^4P^{(1)}(\D_i).
\end{equation}
In a transient configuration, when one hydrogen bond is broken 
before another is formed, it is plausible to assume that the domain is occupied by a number of water molecules different from one. According to 
Eq.~(\ref{eqn: number of HBs}), we find that the average number of HBs formed by a water molecule is $n_{\mathrm{HB}}=3.37$, in 
agreement~\cite{luzar-chandler,science_exp} with the values reported in the literature of $3.3 - 3.6$.

\begin{figure}[h!]
\begin{center}
\includegraphics*[width=.5\textwidth]{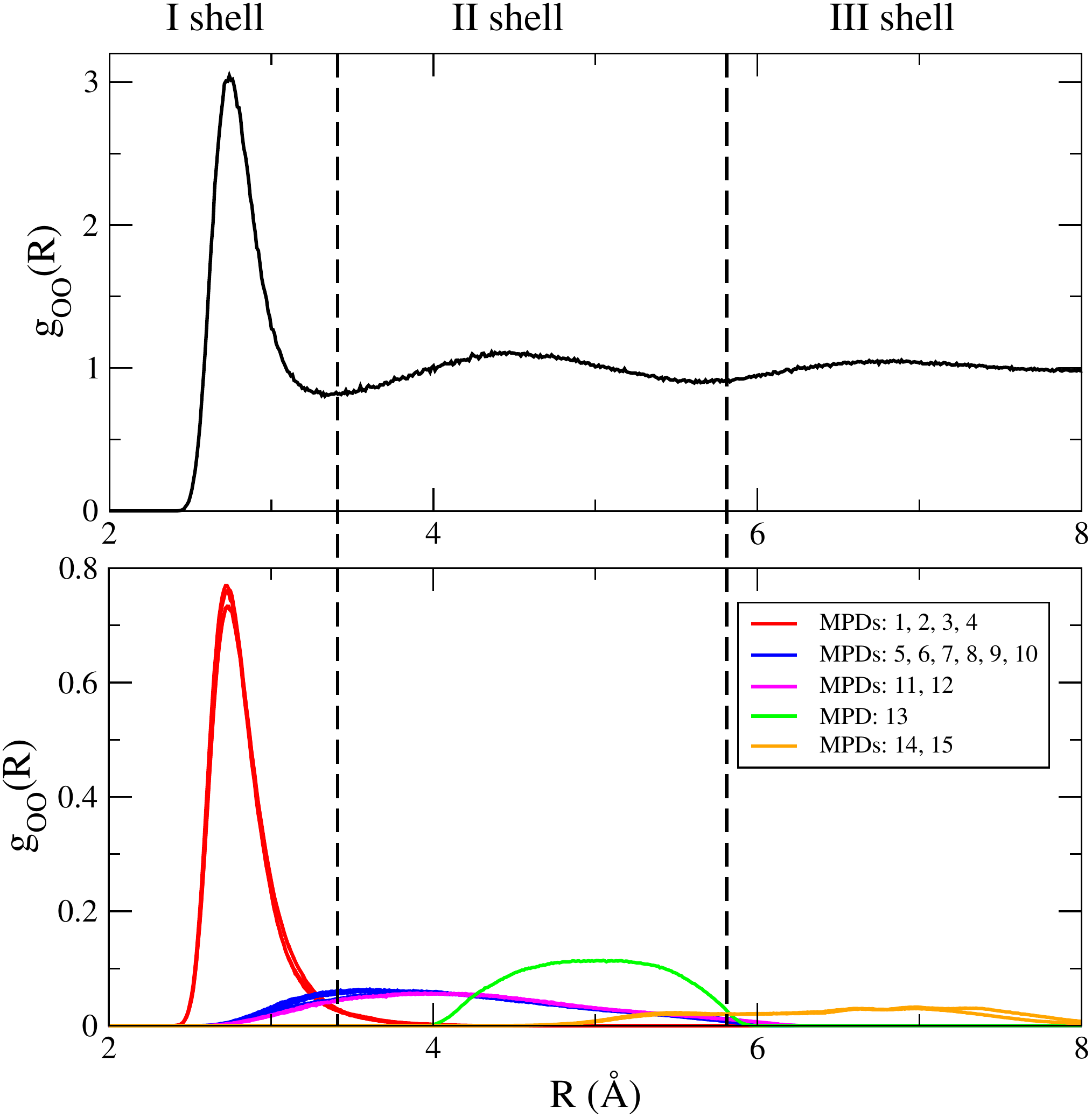}
\caption{Oxygen-oxygen RDFs (colored lines) computed within the MPDs, compared to the total RDF of water at density $\rho_{0}$ (black line). Dashed 
vertical lines are used to show that space partitioning in terms of solvation shells (from the minima of the oxygen-oxygen RDF) or MPDs brings to a 
similar result (for instance the boundary between the second and third solvation shell is not exactly placed at the minimum of the total RDF).}
\label{fig: rdf}
\end{center}
\end{figure}

In the context of the MPDs, solvation shells can be defined by calculating the oxygen-oxygen RDF \textsl{resolved} in each optimized domain. In 
Fig.~\ref{fig: rdf} we show the RDFs computed within the MPDs (lower panel) and compared to the total RDF (upper panel). The $x$-axis has been 
divided in three regions, each representing a solvation shell around the central water molecule. The extent of the first solvation shell is up to about 3.5~\AA. The MPDs labeled from 5 to 12 
(corresponding to the blue and magenta lines in Fig.~\ref{fig: rdf}) only partially occupy the cavities between the domains 1 to 4 of the first 
solvation shell, but they extend to the second solvation shell.
The coordination number in the first solvation shell is 5.1, if computed as the integral of the total RDF up to 3.5~\AA, and 4.5, if computed as the 
sum of the integrals of the partial RDFs. An \textsl{extra-particle}~\cite{sciortino} appears in the first solvation shell, which does not occupy one 
of the four domains of the tetrahedron, but it is delocalized in the interstitial domains 5 to 12. Particularly interesting is the distribution of 
particles in these interstitial domains since the partial RDFs, although very flat, show peaks at about 3.5~\AA. Similar maxima have been 
already observed by Svishchev and Kusalik~\cite{svishchev_structure_1993} from calculation of RDFs resolved in angle. This feature of the RDFs 
suggests the presence of water molecules in non-tetrahedral directions but still penetrating the first solvation shell.



\subsection{Liquid water at {\boldmath$\rho=1.23$ g/cm$^3$}}\label{sec: water 1.23}
The MD trajectory of 150~ps for liquid water at room temperature is generated by employing the TIP4P~\cite{tip4p} model. The system is composed of 
4096 molecules in a cubic box with side length 46.3~\AA.

\begin{figure}[h!]
\begin{center}
\includegraphics*[width=.35\textwidth]{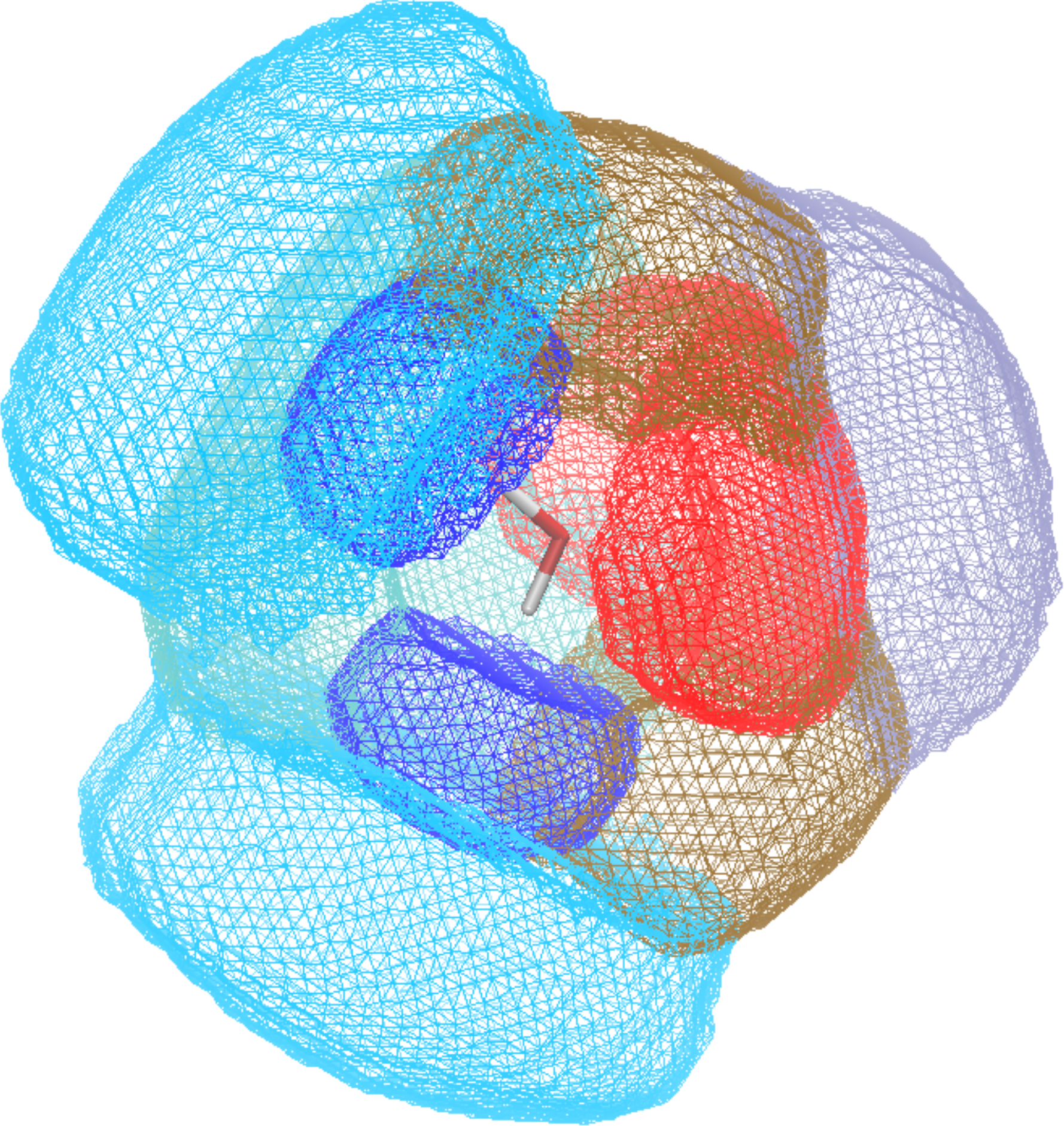}
\includegraphics*[width=.35\textwidth]{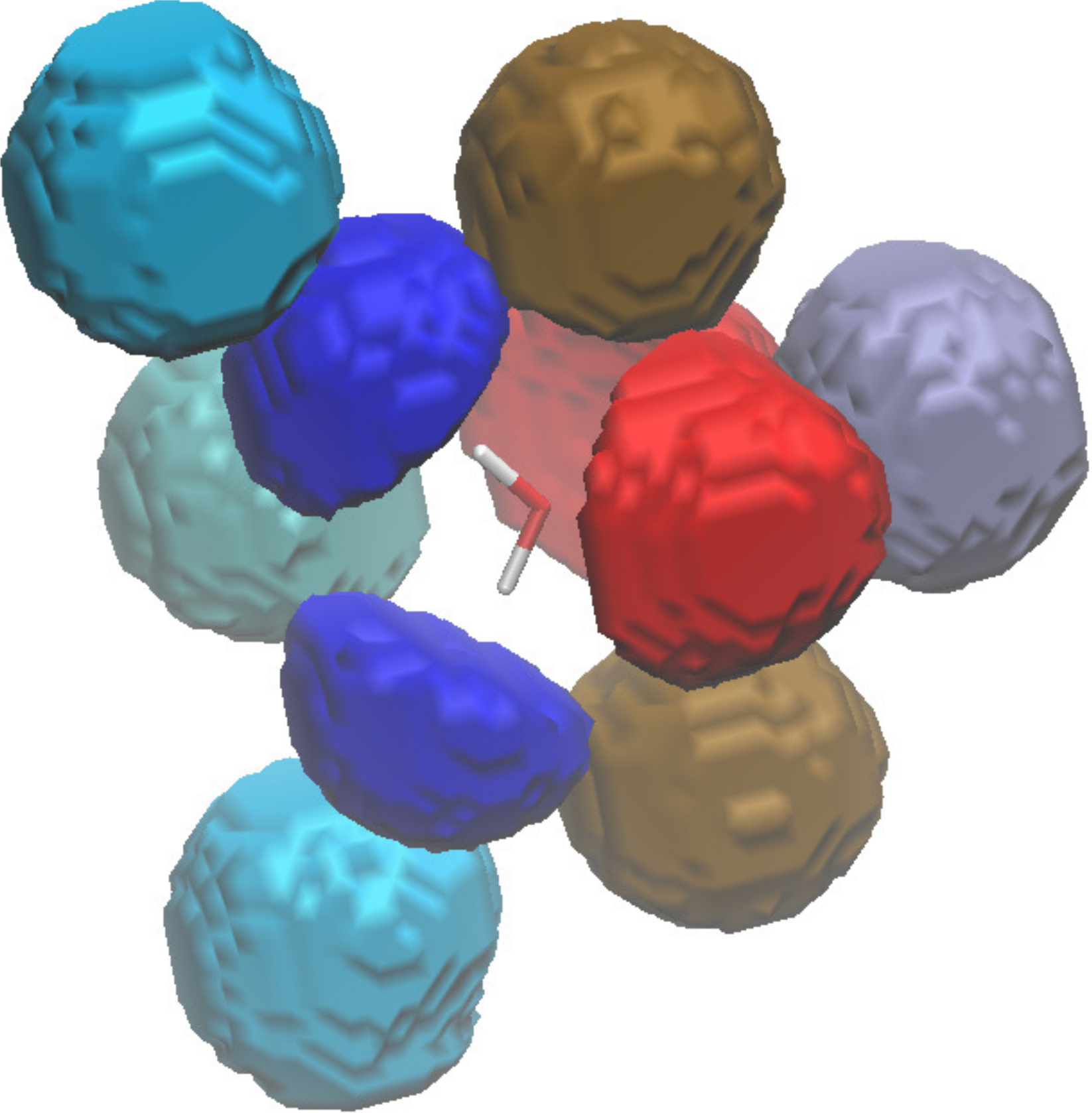}
\caption{Case of water at $\rho=1.23$ g/cm$^3$. Left: Bader domains for water at $\rho=1.23$ g/cm$^3$. Only one domain of those labeled 5 and 6 (one of these two domains is shown as the 
black filled area in Fig.~\ref{fig: observations}) is shown, in order to make the central water molecule visible. Right: MPDs.}
\label{fig: basins h.d.}
\end{center}
\end{figure}

Bader analysis is applied to the two-particle SDF computed for liquid water at the density $\rho=1.23$ g/cm$^{3}$ and 11 domains are identified, in 
contrast to the previous case where 15 domains were defined by Bader analysis. The shapes and positions of the initial domains are shown in 
Fig.~\ref{fig: basins h.d.} (left) along with the MPDs (right). We will present here the results from the MPD analysis in comparison to the 
observations reported in the previous section.

\subsubsection{Comparison with water at $\rho_0$ based on the MPD analysis}
\begin{figure}[h!]
\begin{center}
\includegraphics*[width=.35\textwidth]{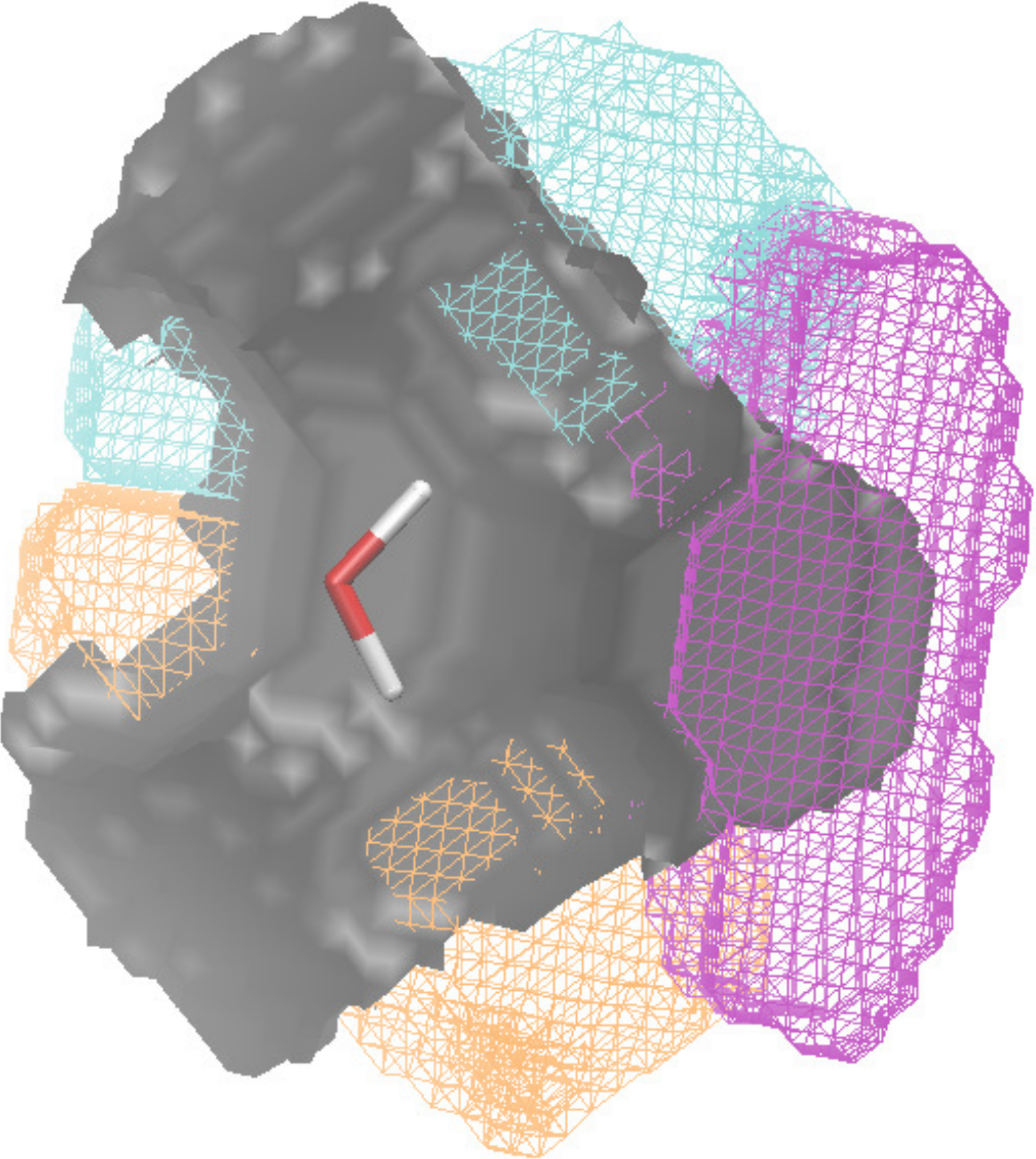}
\caption{Comparison of Bader domains for water at density $\rho_0=1.0$ g/cm$^{3}$ (color grids) and at density $\rho=1.23$ g/cm$^{3}$ (black filled 
areas).}
\label{fig: observations}
\end{center}
\end{figure}

The first solvation shell is unaffected~\cite{saitta,yan} by the increase of density. The remaining domains are arranged closer to the central 
molecule and they are more localized in space.
Since the water molecules are more ``packed'' at higher density and the maxima of the $\rho^{(2)}(\R)$ are less sharp, Bader domains 5 to 10 from Fig.~\ref{fig: basins optimisation 5 to 10} merged into two domains, as shown in Fig.~\ref{fig: observations}.

The MPDs can be grouped according to the corresponding final probability of enclosing only one particle, shown in 
table~\ref{tab: distances h.d.}. If we use again Eq.~(\ref{eqn: number of HBs}) to estimate the value of HBs per molecule, we find 
$n_{\mathrm{HB}}=3.41$. Apart from this aspect, the optimized domains in the first solvation shell are not very much affected by the change of 
density. This observation agrees~\cite{saitta, yan, soper, jedlovszky, straessle} with the literature, namely the MPDs analysis confirms that the 
first solvation shell is quite rigid under density increase.

Looking at the one-particle occupancy probability of the MPDs 5 to 11, listed in table~\ref{tab: distances h.d.}, the optimization procedure seems to 
be more effective in this case of higher density. Also, the average number of particles found in each MPD is very close to one for all domains. This 
effect can be interpreted as a stabilization of the molecules in the second solvation shell due to the packing imposed by the higher density.
\begin{table}
\begin{tabular}{cccccc}
\hline
\hline
Domain label & $\left\langle N\right\rangle_{\textrm{before}} $ & $\left\langle N\right\rangle_{\textrm{after}} $& Volume (\AA$^{3}$) & Distance 
(\AA) 
& $P^{(1)}(\D^*)$\\
\hline
1, 2 & 0.7 & 0.9 & 12.6 & 2.7 & 0.93 \\
3, 4 & 1.7 & 1.0 & 12.0 & 2.8 & 0.78 \\
5, 6 & 5.1 & 1.3 & 14.5& 3.7 & 0.74 \\
7, 8 & 2.7 & 1.3 & 15.0 & 4.0 & 0.72 \\
9 & 2.4 & 1.0 & 15.8 & 4.7 & 0.68 \\
10, 11 & 5.0 & 0.9 & 16.1 & 5.6 & 0.70 \\
\hline
\hline
\end{tabular}
\caption{\label{tab: distances h.d.}List of the MPDs (first column), average number of particles in each volume before (Bader analysis, second 
column) or after the optimization (third column), their volumes (fourth column), distances of their centers of mass from the central oxygen atom 
(fifth column) and the one-particle occupancy probability (sixth column). In the same line of the first column we have put equivalent domains.}
\end{table}

\begin{figure}[h!]
\begin{center}
\includegraphics*[width=.5\textwidth]{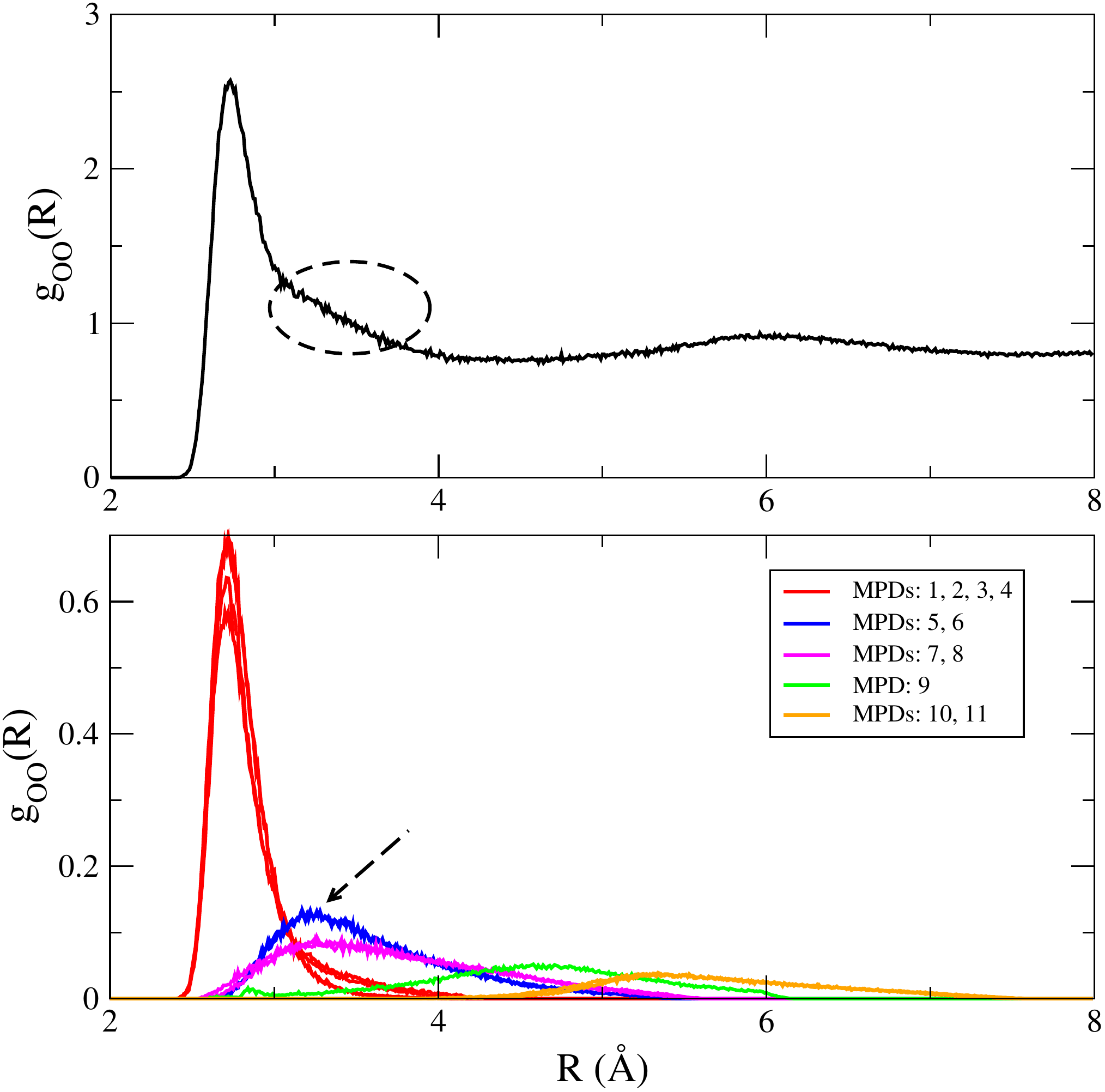}
\caption{Oxygen-oxygen RDFs computed within each MPD, compared to the total RDF of water at density 1.23~g/cm$^{3}$. In the upper panel, the shoulder 
in the highlighted region is produced by the curves in the lower panel which are indicated by the arrow.}
\label{fig: rdf h.d.}
\end{center}
\end{figure}

Comparing the distances of the MPDs from the central molecule at the two densities, $\rho_0$ and $\rho$, we observe a major difference in the behavior of domains 5 and 6, found at a distance of 3.7~\AA~as shown in table~\ref{tab: distances h.d.}, from the behavior of the set of domains at a distance between 4 and 5~\AA~reported in table~\ref{tab: distances}. This result is extremely interesting, as it seems that we have been able to identify, in terms of the MPDs, the location of the 
interstitial oxygen atoms that are mainly affected by the increase of the density. As proven in Fig.~\ref{fig: rdf h.d.}, not all the interstitial 
domains are strongly affected by the change of density (see the difference between the blue and the magenta curves). The pronounced peak (blue line in 
the figure) in the partial RDFs corresponding to the MPDs 5 and 6 at around 3.3~\AA~contributes to the shoulder in the total oxygen-oxygen RDF at the 
same distance (highlighted in the figure by the circle). In the previous analysis we showed that all domains in these regions, namely those labeled 5 
to 12, have similar partial RDFs (blue and magenta lines in Fig.~\ref{fig: rdf}). Instead we observe here that the partial RDFs calculated inside the 
domains 5, 6 show a more pronounced peak than the domains 7, 8. In general, we can observe that the MPDs 5 to 8 enclose the four 
molecules~\cite{saitta} that in non-tetrahedral directions~\cite{jedlovszky, svishchev_structure_1993} mainly contribute to modifications of the 
second solvation shell at increasing density. The hypothesis~\cite{saitta} that, as the density increases, the second solvation shell does not 
continuously collapse on the first shell but interstitial molecules, that are not H-bonded to the first shell molecules, get closer to the central 
water is consistent with our results.

\subsection{$\bf{Na^+}$ in water}\label{sec: sodium in water}
The MD trajectory of 110~ps for a sodium ion in liquid water, at room temperature, is generated by employing the TIP4P~\cite{tip4p} model for water. 
The system is composed of 1024 molecules in a cubic box with side length 31.5~\AA.

The sodium-oxygen SDF is spherically symmetric and it does not contain more information than the sodium-oxygen RDF, therefore the MPDs will have the same spherical distribution around the central ion. The position and shape of the initial domains can be chosen totally 
arbitrarily, since Bader analysis is not efficient in this situation, where the local maxima of the two-particle SDF cannot be properly located 
(maxima of the density are distributed on a sphere, they are not isolated points in 3D space). Also, if the initial domains are chosen within a 
distance of less than about 2.2~\AA, where the RDF (see Fig.~\ref{fig: rdf Na-water}) is very small, the optimization procedure is not 
efficient and the MPDs cannot be identified. This problem is related to the fact that in empty regions, where the probability of finding one particle 
vanishes, a small variation of the region itself does not change this probability, thus fulfilling the optimization condition 
$|P^{(1)}(\D_\tau)-\Prob|\leq\delta$ without an effective modification of the domain. Here, we have considered only domains $\D$ that are enclosed in 
the region of non-zero probability density. We then choose, as initial domains, spheres of different radii, randomly located around the central ion 
and at distances between 2.5~\AA~and 4.5~\AA~from it.

As in the previous applications, we will first show the MPDs and then we will introduce some analysis tools to determine the properties of the MPDs.

\subsubsection{Analysis of the MPDs}
\begin{figure}
\begin{center}
\includegraphics*[width=.45\textwidth]{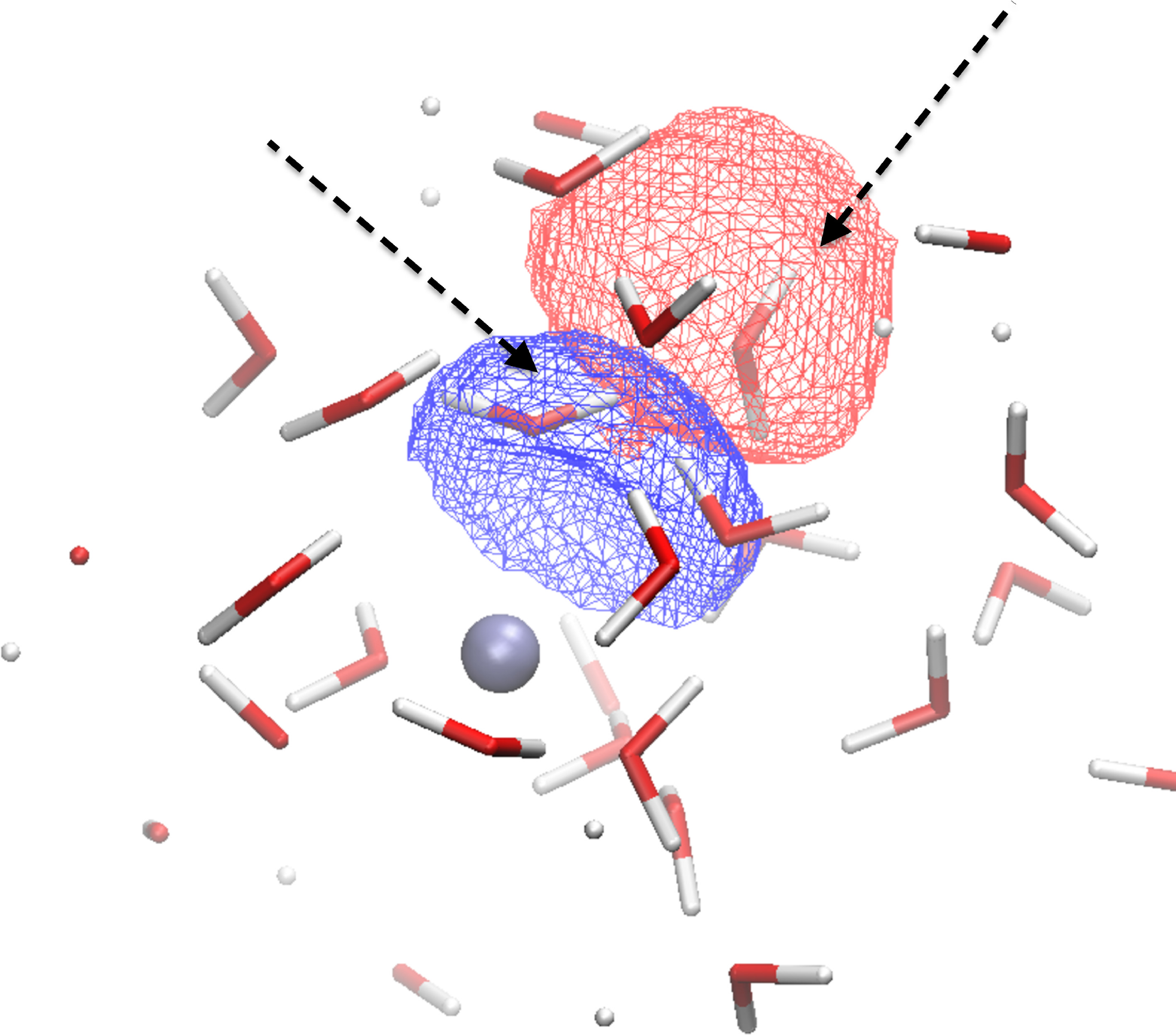}
\caption{Case of sodium diluted in water. MPDs in the first and second solvation shells of Na$^+$. From a snapshot of the trajectory, some water molecules in the vicinity of the 
central ion are also shown. We see that each domain is occupied by only one molecule (indicated by the arrows).}
\label{fig: domain Na-water}
\end{center}
\end{figure}

The optimization procedure identifies two sets of MPDs, associated to the first and second solvation shells of water around the central sodium ion. 
They are shown in Fig.~\ref{fig: domain Na-water} as blue and red grid volumes, with distances from the central ion of 2.3 and 4.4~\AA, respectively. 
These values are listed in table~\ref{tab: distances Na}, along with the other properties associated to the two MPDs.

\begin{table}
\begin{tabular}{cccccc}
\hline
\hline
Shell & Volume (\AA$^{3}$)& Distance (\AA) & $\Omega$ (deg$^2$) & $n_{\Omega}$ & $P^{(1)}(\D^*)$\\
\hline
I & 12.0 & 2.3 & 135.6 & 5.3 (5.9) & 0.78 \\
II & 16.2 & 4.4 & 39.9 & 18.1 (19.3) & 0.62\\
\hline
\hline
\end{tabular}
\caption{\label{tab: distances Na}List of properties of the MPDs calculated for Na$^+$ in water. In the first column we list the position of the 
domain, in one of the two solvation shells around the sodium ion, in the second we calculate the volumes occupied by the domain, in the third column 
we report the distance of the center of mass of the domain from the central ion, in the fourth column we show the solid angle occupied by the domain 
and in the fifth column, the coordination number associated to it. In parenthesis, we compare the value of the coordination number determined by 
integrating the RDF up to $R=$3.3,5.6~\AA. In the sixth column the values of the one-particle occupancy probability are given.}\end{table}
\begin{figure}
\begin{center}
\includegraphics*[width=.5\textwidth]{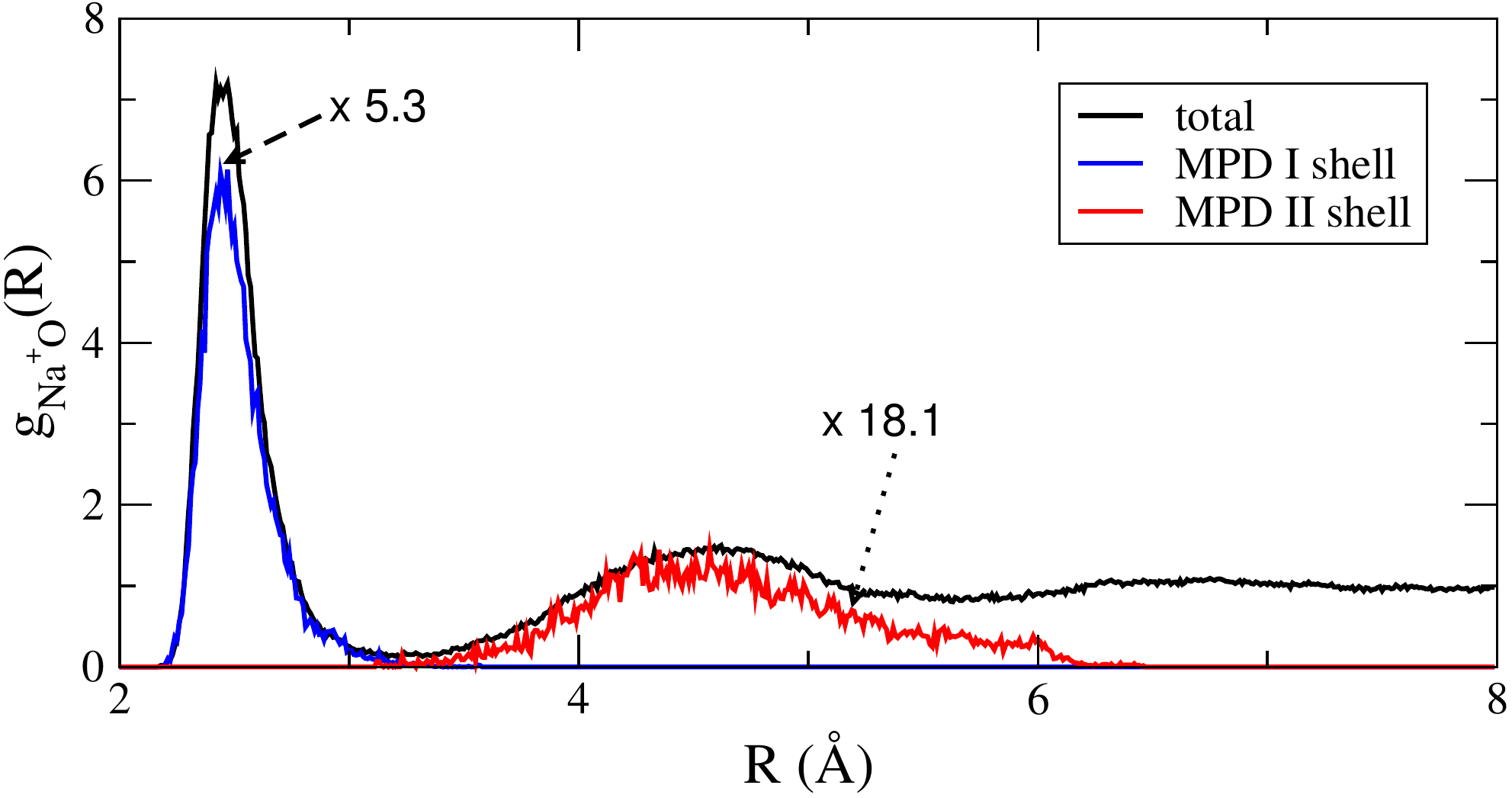}
\caption{Total and partial (calculation restricted to each given domain) ion-oxygen RDF. The blue and red curves are determined by calculating the 
RDF in the domains shown in Fig.~\ref{fig: domain Na-water} and by multiplying it by the corresponding $n_{\Omega}$ (from table~\ref{tab: distances 
Na}).}
\label{fig: rdf Na-water}
\end{center}
\end{figure}

The coordination number $n_{\Omega}$ in the two shells is determined as the ratio of the solid angle occupied by the domain $\Omega$ and the total solid angle, whose values are shown in table~\ref{tab: distances Na}. In particular, the coordination 
numbers $n_{\Omega}$ are in very good agreement with those determined by integrating the RDF up to the distances 3.3 and 5.6~\AA, for the first and 
second solvation shell, respectively. As also observed in the case of water, slightly smaller values are determined by our analysis, if compared to 
the integral of the RDF. The total RDF can be reconstructed by determining its value inside the MPDs. The comparison between the total 
RDF and the partial ones (in each domain) is shown in Fig.~\ref{fig: rdf Na-water}. The curves calculated in the two domains are multiplied by the 
corresponding $n_{\Omega}$ in order to obtain the contribution from the whole shell.

\section{Conclusions}\label{sec: conclusion}
We extended the study of the structure of a liquid system by the identification of MPDs, borrowing the idea from what is now routinely done in electronic structure analysis, and we developed a first application to few case studies. We derived the formalism necessary to define the probability of finding one and only one particle in a certain region of space $\D$ and we developed an algorithm for the optimization of this region under the request that the probability is maximized. The geometric optimization is formulated in terms of shape derivatives, thus allowing to use the LSM to solve the optimization problem.

The case study obtained the features of the MPDs for water at different densities and for a solvated sodium ion in water. A qualitative analysis of the domains in water has allowed us to describe the modification of the second solvation shell by increasing pressure. More quantitative observations have also illustrated the presence of water molecules in the interstitial spaces of the first solvation shell, that are not H-bonded to the central molecule. These molecules contribute to the modification of the second solvation shell of water at increasing density. Looking at the RDFs restricted to the MPDs, we have been able to indirectly achieve a 3D resolution that is totally lacking in the (standard) full oxygen-oxygen RDF, since we know exactly the spatial organization of the MPDs. Therefore, this result proves that this new approach adds to standard techniques, providing structural 3D information on liquids. In the case of a sodium ion in water, we reconstructed the sodium-oxygen RDF from the partial contributions evaluated inside the optimized domains and we determined the coordination numbers associated to these domains. We used a criterion to identify the solvation shells, based on the identification of regions where the partial RDFs, calculated within the MPDs, are close to zero, rather than on the identification of the minima of the total RDF. Notice that, even though we started the research for the MPDs around the sodium ion from the information obtained from the RDFs (but only as input in the analysis, the distance from the central ion of the maxima of the sodium-oxygen RDF), we have extracted the full 3D shape of the domains containing the oxygen atoms around the sodium ion. In this case, in fact the sodium-oxygen SDF cannot provide this information, since the problem has spherical symmetry.

We investigated the properties of the MPDs that are determined from the maximization of the one-particle occupancy probability. Along similar lines 
one could analyze probabilities associated to more than one particle or focus on other atoms (hydrogens, for instance). This could be used to give a 
many-body definition of solvation shells.

The proposed method results to be an efficient tool to complement the standard analysis techniques used in the study of the structure of a liquid 
system and to give the 3D image of the organization of the space around a given molecule, atom or ion. To conclude, let us say that, despite the somehow elaborate 
mathematical construction presented in the paper, the physical interpretation of the MPD approach is simple. It allows to identify at 3D level the statistical meaning of the positions of 
particles surrounding a given central molecule or ion. To do that, it looks at new and different (from those in standard use) probabilities, allowing to better \textsl{isolate} the 3D statistical arrangement of the observed particles.

\section*{Acknowledgements}
F.~A. and R.~V. acknowledge the support received from the European Science Foundation (ESF) for the activity entitled ``Molecular Simulations in Biosystems and Material Science''. G.~C. is grateful to E.~K.~U. Gross for his hospitality at the Max-Planck-Institute in Halle and his support all along the long gestation of this work. The authors thank D. Borgis for fruitful discussions at the initial stages of this work.

\appendix

\section{Probability density}\label{app: probability}
We have illustrated how to use the MPDs to investigate the local structure of a liquid around a given center. In the examples discussed in the 
paper, the center was chosen as a water molecule, in the case of liquid water, and as the sodium ion, in the case where we analyze the structure 
of the water solvation shells around this ion. Therefore, the probability density $\dens$ introduced in Eq.~(\ref{eqn: def probability}) has to be identified as 
the conditional probability density of the configuration $\R^N$ of the $N$ oxygen atoms, given a certain configuration $\lbrace\R_0,\R^{\mathrm 
H_1}_0,\R^{\mathrm H_2}_0\rbrace$ of the central water molecule,
\begin{equation}\label{eqn: dens as conditional probability}
\dens= \rho\left(\R^N\left|\right.\R_0,\R^{\mathrm H_1}_0,\R^{\mathrm H_2}_0\right),
\end{equation}
or of the central sodium ion $\lbrace\R^{\mathrm{Na}^+}_0\rbrace$, 
\begin{equation}
\dens= \rho\left(\R^N\left|\right.\R^{\mathrm{Na}^+}_0\right).
\end{equation}
We explicitly discuss below only the case of water, while the expressions for sodium in water are obtained by replacing 
$\lbrace\R^{\mathrm{Na}^+}_0\rbrace$ to $\lbrace\R_0,\R^{\mathrm H_1}_0,\R^{\mathrm H_2}_0\rbrace$.

Here we label with ``0'' the positions of the oxygen atom and the two hydrogen atoms of the central molecule. To be more explicit on the relation 
between the density in Eq.~(\ref{eqn: dens as conditional probability}) and the standard configurational canonical density $\rho^{can}(\R_0,\R^N,\R^{\mathrm H_1}_0,\R^{\mathrm H_2}_0,{\R^{\mathrm 
H_1}}^N,{\R^{\mathrm H_2}}^N)$, that is a function of all atomic positions, i.e. oxygens and hydrogens, we give the following expression of $\dens$,
\begin{align}
\dens =&  \int d{\R^{\mathrm H_1}}^N \, d{\R^{\mathrm H_2}}^N\,
\frac{\rho^{can}\left(\R_0,\R^N,\R^{\mathrm H_1}_0,\R^{\mathrm H_2}_0,{\R^{\mathrm 
H_1}}^N_1,{\R^{\mathrm H_2}}^N\right)}{P_m\left(\R_0,\R^{\mathrm H_1}_0,\R^{\mathrm H_2}_0\right)}.\label{eqn: conditional probability}
\end{align}
Since $\dens$ is a conditional probability density, the denominator represents the marginal probability of finding the central molecule in the 
configuration $\lbrace\R_0,\R^{\mathrm H_1}_0,\R^{\mathrm H_2}_0\rbrace$, namely
\begin{align}\label{eqn: marginal probability}
P_m\left(\R_0,\R^{\mathrm H_1}_0,\R^{\mathrm H_2}_0\right) = \int d\R^N \, d{\R^{\mathrm H_1}}^N \, d{\R^{\mathrm 
H_2}}^N\,
\rho^{can}\left(\R_0,\R^N,\R^{\mathrm H_1}_0,\R^{\mathrm H_2}_0,{\R^{\mathrm H_1}}^N,{\R^{\mathrm H_2}}^N\right).
\end{align}

\section{The level set method}\label{app: lsm}
We present here in detail the LSM, defining the shape derivative of $\Prob$ and introducing the level set function.

\subsection{Shape derivative}\label{sec: shape}
$\Prob$ belongs to a specific class of what is normally called a set function. We now introduce few analysis tools used when treating in general set 
functions.

The shape derivative of a set function of the form
\begin{equation}\label{eqn: general set function}
F(\D)=\int_{\D}d\R \,f(\R)
\end{equation}
is defined as the variation with respect to the integration domain $\D$. In other words $F(\D)$ can be written in terms of a ``density'', $f(\R)$, 
and, in the following section, we give the explicit definition of the $f(\R)$, interesting us, from Eq.~(\ref{eqn: prob with characteristic 
function}). 

If the deformation law of $\D$ is given as
\begin{equation}\label{eqn: deformation law}
\D\rightarrow \D_{\tau} = \left\lbrace \R_{\tau}=\R+\V d\tau \,|\,\R\in \partial\D \right\rbrace
\end{equation}
where $d\tau$ is a fictitious time step and $\V$ is a velocity field (to be specified below), the time derivative of $F(\D)$ can be calculated according to
\begin{equation}\label{eqn: derivative of F}
\frac{d F(\D)}{d\tau} =\lim_{d\tau\rightarrow 0}\frac{1}{d\tau} \left[\int_{\D_{\tau}} d\R_{\tau}\,f(\R_{\tau}) - \int_{\D}d\R\,f(\R)\right].
\end{equation}
Performing, in the first term in square brackets, the change of variable
\begin{align}\label{eqn: new variables}
\R_{\tau} = \R +\V d\tau, \quad
d\R_{\tau}=d\R\left(1+\nabla\cdot\V d\tau+\mathcal O(d\tau^2)\right)
\end{align}
and expanding, for small time increment $d\tau$, $f(\R_{\tau})$ around $\R$
\begin{equation}\label{eqn: expansion of f}
f(\R_{\tau}) = f(\R) + \V\cdot \nabla f(\R) d\tau,
\end{equation}
Eq.~(\ref{eqn: derivative of F}) becomes
\begin{eqnarray}
\frac{d F(\D)}{d\tau} &=& \int_{\D} d\R\, \nabla\cdot\left(\V f(\R)\right) \label{eqn: divergence theorem}\\
&=& \int_{\partial\D} ds\, \mathbf n(\R)\cdot\V f(\R)\equiv\shape F(\D) \label{eqn: shape derivative}
\end{eqnarray}
where the divergence theorem has been used to derive Eq.~(\ref{eqn: shape derivative}) from Eq.~(\ref{eqn: divergence theorem}). In the last line $ds$ 
is the surface element of the boundary $\partial \D$ of $\D$ and $\mathbf n(\R)$ is the unit vector normal to the surface at $\R$. Eq.~(\ref{eqn: 
shape derivative}) defines the \textsl{shape derivative}, indicated by the symbol $\shape$, of $F(\D)$ and expresses the variation of $F(\D)$ with 
respect to the variation of $\D$. The change in the domain $\D$ is expressed in terms of a global deformation of the boundary $\partial \D$, which, in 
turn, is determined by the velocity field $\V$. For our purpose, that is the maximization of $F(\D)$, we can choose the field $\V$ such that $\shape 
F(\D)\geq 0$ and follow the evolution of such a field up to find the final $\shape F(\D)=0$. The condition $\shape F(\D)\geq 0$ is automatically 
satisfied if we choose the velocity field as 
\begin{equation}\label{eqn: velocity field 1}
\V=\mathbf n(\R) f(\R).
\end{equation}

\subsection{Shape derivative for the MPDs}
The comparison between Eqs.~(\ref{eqn: def probability}) and~(\ref{eqn: general set function}) shows that when we express the probability $\Prob$ in 
terms of a density function similar to $f(\R)$, such density depends on $\D_c$, i.e. on $\D$. Therefore, when calculating the variations of $F(\D)$ 
from Eq.~(\ref{eqn: general set function}) with respect to $\D$, we need to include also variations of $f(\R)$ with respect to $\D_c$, i.e. $\D$. In 
our case, then, the transformation in Eq.~(\ref{eqn: new variables}), restricting ourselves to the case $\mathbf v_j(\R)=\mathbf v(\R_j)$, becomes
\begin{align}
\R_{j,\tau}&=\R_j+\mathbf v(\R_j)d\tau \quad\mbox{with }j=1,\ldots,N\label{eqn: evolved Rj}
\end{align}
and the volume elements are
\begin{align}
d\R_{j,\tau}&=d\R_j\left(1+\nabla_{\R_j}\cdot\mathbf v (\R_j)d\tau\right). \label{eqn: evolved dRj}
\end{align}
The time derivative of the probability $\Prob$ under the variation $\D\rightarrow \D_{\tau}$ is
\begin{align}
\frac{d\Prob}{d\tau}&=\nonumber \\
\lim_{d\tau\rightarrow0}&\frac{1}{d\tau}\sum_{i=1}^N\Big[\int_{\D_{\tau}}d\R_{i,\tau}\left(\prod_{l\neq 
i}^N\int_{\D_c,\tau}d\R_{l,\tau}\right)\rho(\R_{\tau}^N)- 
\int_{\D}d\R_{i}\left(\prod_{l\neq i}^N\int_{\D_c}d\R_{l}\right)\rho(\R^N)\Big]
\label{eqn: dP/dtau 1}
\end{align}
where $\D_{c,\tau}$ is the complementary volume to $\D_\tau$, $d\R_\tau^{N-1}=d\R_{2,\tau}\ldots d\R_{N,\tau}$ (similarly for $d\R^{N-1}$) and 
$\R^N_\tau=\R_{1,\tau},\ldots ,\R_{2,\tau}$. Using Eq.~(\ref{eqn: evolved dRj}) and the expansion of the density $\rho(\R_{\tau}^N)$ up to linear 
order (see Eq.~(\ref{eqn: expansion of f})) in the displacement from $\R^N$,
\begin{align}
\rho(\R_{\tau}^N)=\rho(\R^N)+\sum_{j=1}^N\left[\nabla_{\R_j}\rho(\R^N)\right]\cdot \mathbf v(\R_j)d\tau,
\end{align}
the shape derivative of $\Prob$ becomes
\begin{align}
\frac{d\Prob}{d\tau}=
\sum_{i=1}^N\int_{\D}&d\R_{i}\left(\prod_{l\neq i}^N\int_{\D_c}d\R_{l}\right)
 \Bigg[\nabla_{\R_i}\cdot\Big(\rho(\R^N)\mathbf v(\R_i)\Big)+ 
 \sum_{j\neq i}^N\nabla_{\R_j}\cdot\Big(\rho(\R^N)\mathbf v(\R_j)\Big)\Bigg].
\label{eqn: dP/dtau 2}
\end{align}
The first term on the right-hand-side can be treated exactly as we have done in Eq.~(\ref{eqn: divergence theorem}), leading to
\begin{align}
\int_{\D}d\R_{i}\left(\prod_{l\neq i}^N\int_{\D_c}d\R_{l}\right)\nabla_{\R_i}\cdot\Big(\rho(\R^N)\mathbf v(\R_i)\Big)= 
\int_{\partial\D}ds_{i}\,\mathbf n(\R_i)\cdot\mathbf v(\R_i)\left(\prod_{l\neq i}^N\int_{\D_c}d\R_{l}\right)\rho(\R^N)\nonumber\\
=\int_{\partial\D}ds\,\mathbf n(\R)\cdot\mathbf v(\R)\int d\R_i\,\delta\left(\R_i-\R\right)\left(\prod_{l\neq 
i}^N\int_{\D_c}d\R_{l}\right)\rho(\R^N).\label{eqn: first term in dP/dtau}
\end{align}
The second line is obtained from the first by introducing a $\delta$-function in order to change the variable $\R_i$ to $\R$ in the surface integral. 
Notice now that the integral over the variable $\R_i$ is performed over the whole space.

The second term in Eq.~(\ref{eqn: dP/dtau 2}) has still to be appropriately simplified. The $N-1$ integrals of the second term in square brackets are 
transformed, one by one, in $N-1$ surface integrals over $\partial\D_c$ with $\nabla_{\R_j}\cdot$ replaced by the normal vector to the surface, 
$\mathbf n_c(\R_j)=-\mathbf n(\R_j)$ (when the volume $\D$ changes in the direction indicated by $\mathbf n$, the volume $\D_c$ changes in the 
direction $-\mathbf n$, since the boundaries of $\D$ and $\D_c$ are the same, being them complementary volumes to each other). Therefore, we write 
explicitly each term of the sum over the index $j$ in Eq.~(\ref{eqn: dP/dtau 2})
\begin{align}
\int_{\D}d\R_{i}\left(\prod_{l\neq i}^N\int_{\D_c}d\R_{l}\right)&\sum_{j\neq i}^N\nabla_{\R_j}\cdot\Big(\rho(\R^N)\mathbf v(\R_j)\Big)= \nonumber \\
&\int_{\D}d\R_{i}\int_{\partial\D_c}ds_1\left(\prod_{l\neq i,1}\int_{\D_c}d\R_{l}\right)\mathbf n_c(\R_1)\cdot\mathbf v(\R_1) \rho(\R^N) \nonumber \\
&+\int_{\D}d\R_{i}\int_{\partial\D_c}ds_2\left(\prod_{l\neq i,2}\int_{\D_c}d\R_{l}\right)\mathbf n_c(\R_2)\cdot\mathbf v(\R_2) \rho(\R^N) +\ldots
\end{align}
obtaining
\begin{align}
\int_{\D}d\R_{i}\left(\prod_{l\neq i}^N\int_{\D_c}d\R_{l}\right)&\sum_{j\neq i}^N\nabla_{\R_j}\cdot\Big(\rho(\R^N)\mathbf v(\R_j)\Big)= \nonumber \\
&\int_{\D}d\R_{i}\sum_{j\neq i}^N\int_{\partial\D_c}ds_j\left(\prod_{l\neq i,j}\int_{\D_c}d\R_{l}\right)\mathbf n_c(\R_j)\cdot\mathbf 
v(\R_j)\rho(\R^N).
\end{align}
We introduce now a $\delta$-function, in order to make a change of variable $\R_j\rightarrow \R$ and to write the above integral in compact form, 
namely
\begin{widetext}
\begin{align}
\int_{\D}d\R_{i}&\left(\prod_{l\neq i}^N\int_{\D_c}d\R_{l}\right)\sum_{j\neq i}^N\nabla_{\R_j}\cdot\Big(\rho(\R^N)\mathbf v(\R_j)\Big)= \nonumber \\
&\int_{\D}d\R_{i}\sum_{j\neq i}^N\int_{\partial\D_c}ds\,\mathbf n_c(\R)\cdot\mathbf v(\R)\int d\R_j\,\delta\left(\R_j-\R\right)\left(\prod_{l\neq 
i,j}\int_{\D_c}d\R_{l}\right)\rho(\R^N) \label{eqn: from nc}\\
&=-\int_{\D}d\R_{i}\sum_{j\neq i}^N\int_{\partial\D}ds\,\mathbf n(\R)\cdot\mathbf v(\R)\int d\R_j\,\delta\left(\R_j-\R\right)\left(\prod_{l\neq 
i,j}\int_{\D_c}d\R_{l}\right)\rho(\R^N).\label{eqn: to n}
\end{align}
\end{widetext}
From Eq.~(\ref{eqn: from nc}) we have obtained Eq.~(\ref{eqn: to n}) by replacing $\mathbf n_c$ with $-\mathbf n$ and by using the property that the 
boundary of $\D_c$ is the boundary of $\D$. As previously observed for Eq.~(\ref{eqn: first term in dP/dtau}), the integral over the variable $\R_j$ 
is performed over the whole space.

Combining the results in Eqs.~(\ref{eqn: first term in dP/dtau}) and~(\ref{eqn: to n}), we derive the expression for the shape derivative of the 
one-particle occupancy probability as
\begin{widetext}
\begin{align}
\frac{d\Prob}{d\tau}=
\int_{\partial\D}ds\,\mathbf n(\R)\cdot\mathbf v(\R)\,\left\lbrace \sum_{i=1}^N\Bigg[\int d\R_i\,\delta\left(\R_i-\R\right)\left(\prod_{l\neq 
i}^N\int_{\D_c}d\R_{l}\right)\right.\\
\left.-\int_{\D}d\R_{i}\sum_{j\neq i}^N\int d\R_j\,\delta\left(\R_j-\R\right)\left(\prod_{l\neq 
i,j}\int_{\D_c}d\R_{l}\right)\Bigg]\rho(\R^N)\right\rbrace.
\end{align}
\end{widetext}
The characteristic functions $\Upsilon_{\D}(\R_j)$ and $\Upsilon_{\D_c}(\R_j)=1-\Upsilon_{\D}(\R_j)$ $\forall j$ are used in order to extend the 
integrals over $\D$ and $\D_c$ to the whole space. We then obtain
\begin{align}
\frac{d\Prob}{d\tau}=\int_{\partial\D}ds\,\mathbf n(\R)\cdot\mathbf v(\R)f_\D(\R)\label{eqn: dP/dtau 4}
\end{align}
with
\begin{align}
f_\D(\R) =&\left\langle \sum_{i=1}^N\Bigg[\delta\left(\R_i-\R\right)\prod_{l\neq i}^N\left(1-\Upsilon_{\D}(\R_l)\right)-
\sum_{j\neq 
i}^N\Upsilon_{\D}(\R_i)\delta\left(\R_j-\R\right)\prod_{l\neq i,j}^N\left(1-\Upsilon_{\D}(\R_l)\right)\Bigg]\right\rangle.\label{eqn: f_D}
\end{align}
If the arbitrary velocity field in Eq.~(\ref{eqn: dP/dtau 4}) is chosen to be $\mathbf v(\R)=\mathbf n(\R) f_{\D}(\R)$ as in Eq.~(\ref{eqn: velocity 
field 1}), the condition $\shape\Prob=d\Prob/d\tau \geq 0$ is again automatically satisfied. To evaluate $\widetilde{\Gamma}_{\D}^{(i)}(\R;\R^{N})$, we proceed as follows:
\begin{itemize}
\item First term on the right-hand-side of Eq.~(\ref{eqn: f_D}): if $\R_i=\R$ and all the other $N-1$ particles are in $\D_c$, 
then the term is 1, irrespective of whether $\R$ is in $\D$ or in $\D_c$; otherwise it is 0.
\item Second term on the right-hand-side of Eq.~(\ref{eqn: f_D}): if $\exists\, k\neq i$ such that $\R_k=\R$, while $\R_i\in 
\D$ and the other $N-2$ $R_j\in\D_c$, then the term is 1, irrespective of whether $\R$ is in $\D$ or in $\D_c$; otherwise it is 0.
\end{itemize}

\subsection{Level set function}\label{sec: level set function}
In order to regularize the mathematical treatment of the characteristic functions, it is useful to introduce a family of regular functions 
collectively defined as level set functions. A level set function $\lsf$ is defined by the property
\begin{equation}
\lsf\left\lbrace
\begin{array}{ll}
<0, & \R\in\D \\
=0, & \R\in\partial\D \\
>0, & \R\not\in\D
\end{array}
\right. \quad \forall\,\tau,
\end{equation}
such that $\partial\D$ is identified as the set of zeros of $\lsf$. The normal vector to the iso-surfaces of this smooth function is defined as
\begin{equation}\label{eqn: normal vector}
\mathbf n(\R) = \frac{\nabla\lsf}{\left|\nabla\lsf\right|}
\end{equation}
and, in particular for the iso-surface $\lsf=0$, $\mathbf n(\R)$ is the normal vector to the boundary of $\D$ that appears in Eq.~(\ref{eqn: dP/dtau 
4}).

To define the evolution of the set, we ask that the level set function $\lsf$ be such that its total time-derivative is 0. This results in the 
following evolution equation
\begin{equation}\label{eqn: lagrangian derivative of lsf 2}
0=\frac{d\phi(\R_\tau,\tau)}{d\tau} = \frac{\partial\phi(\R_\tau,\tau)}{\partial \tau}+\mathbf v(\R_\tau)\cdot\nabla\phi(\R_\tau,\tau),
\end{equation}
where the velocity field $\V$ at $\R_\tau$ will be chosen to be
\begin{equation}
\mathbf v(\R_\tau) = \mathbf n(\R_\tau) f_{\D}(\R_\tau) = \frac{\nabla\phi(\R_\tau,\tau)}{\left|\nabla\phi(\R_\tau,\tau)\right|}f_{\D}(\R_\tau).
\end{equation} 
Eq.~(\ref{eqn: lagrangian derivative of lsf 2}) guarantees that the iso-surface $\phi(\R_\tau,\tau)=0$ mimics the evolution of the boundary of $\D$ 
according to the deformation law determined by imposing the condition $\shape\Prob\geq 0$ on the shape derivative of the one-particle occupancy 
probability. It is important to underline that imposing Eq.~(\ref{eqn: lagrangian derivative of lsf 2}) means that, at the boundary of $\D$, the value 
of the level set function does not change in time. Therefore, we are able to identify at all times the set of points defining $\partial\D_\tau$. 
Further, by imposing that Eq.~(\ref{eqn: lagrangian derivative of lsf 2}) is valid everywhere in space, we are obtaining the deformation law 
$\D\rightarrow \D_\tau$ by evolving the auxiliary level set function. The advantages are the possibility of calculating the normal vector $\mathbf 
n(\R)$, being $\lsf$ a smooth function of $\R$ by construction, and of being able to identify at each time the domain $\D_\tau$ (the characteristic 
function is constructed by knowing where $\lsf$ is positive). We approximate the right-hand-side of Eq.~(\ref{eqn: lagrangian derivative of lsf 2}) by 
a finite difference
\begin{align}
\phi(\R_\tau,\tau+d\tau) =&\phi(\R_\tau,\tau) - d\tau f_\D(\R_\tau) \left|\nabla\phi(\R_\tau,\tau)\right|,\label{eqn: evolution lsf}
\end{align}
using Eq.~(\ref{eqn: normal vector}) for the unitary vector normal to the boundary of $\D$ (or to the iso-surface $\phi(\R_\tau,\tau)=0$). This 
equation is then transformed in an algorithm that determines the evolution of the domains $\D_i$, given an initial condition for the level set 
function corresponding to the choice of initial domains discussed in the text.

\section{The ideal gas}\label{app: ideal gas}
\newcommand{\Pnv}{P^{(n)}(v)}
The probability $\Prob$ can be analytically evaluated for the ideal gas. For a non-interacting system, $\Prob$ does not depend on the shape of the 
domain $\Delta$ but only on its volume $v$. Thus, there is total degeneracy in the shape when looking for an optimal domain because only the volume 
(and not the shape) of the MPD can be determined from the maximization of the one-particle occupancy probability. This fact will reduce the 
calculation of the MPD to a trivial analysis exercise which we will give below. 

Eq.~(\ref{eqn: def probability}) in this case becomes
\begin{align}\label{eqn: Pnv IG}
\Prob =& \frac{1}{V^N}\binom{N}{1}\int_v d\R^1\int_{V-v} d\R^{N-1}
=N\left(\frac{v}{V}\right)\left(1-\frac{v}{V}\right)^{N-1}
\end{align}
where $V^N$ is the configurational partition function for the ideal gas. Here we have indicated the complementary volume of $v$ as $V-v$. The 
thermodynamic limit is now easily obtained by observing that
\begin{align}
\lim_{N\rightarrow \infty}\Prob = \left(\frac{v}{v_0}\right)
\lim_{N\rightarrow \infty}\left(1-\frac{v}{v_0}\frac{1}{N}\right)^{N-1} = \frac{v}{v_0}e^{-\frac{v}{v_0}}
\label{eqn: limit 1}
\end{align}
where $v_0=V/N$ is kept constant and we have eliminated any dependence on $V$. It is natural to ask at this point what is the volume $v$ that 
maximizes this probability and what is the value of this maximum probability. We can answer these questions by just differentiating Eq.~(\ref{eqn: 
limit 1}) with respect to $v$, since in this case the shape derivative reduces to the standard derivative. We find that $dP^{(1)}(v)/dv=0$ if $v=v_0$. 
For the value of the probability at the optimal volume, we find $P^{(1)}(v_0)=e^{-1}=0.37$. 

The values of the one-particle occupancy probability for the MPDs and the volumes of the MPDs have been shown in the main text. We observe in the 
three examples that the one-particle occupancy probability in the absence of interactions is much smaller than those calculated for water. The 
interactions among the particles of the liquid indeed stabilize their distribution and it is in particular the repulsion that prevents the occupation 
of a MPD by more than one particle (that would lead to the decrease of the value of the probability). Moreover, the average volumes per particle $v_0$ 
calculated for water at $\rho_0$ is 30.0~\AA$^3$, for water at $\rho=1.23\rho_0$ is 24.2~\AA$^3$ and for sodium diluted in water is 30.5~\AA$^3$. 
These 
values represent also the volumes of the MPDs in the ideal gas at the same densities. The volumes predicted for the ideal gas are larger than all 
those calculated for water in the examples, suggesting that interactions make the particles less mobile than what is expected in a non-interacting 
situation.

%

\end{document}